\documentclass[conference]{IEEEtran}
\IEEEoverridecommandlockouts

\usepackage{cite}
\usepackage{amsmath,amssymb,amsfonts}
\usepackage{graphicx}
\usepackage{textcomp}
\usepackage{xcolor}

\usepackage{booktabs}
\usepackage{lipsum}
\usepackage{caption}
\usepackage{subcaption}
\usepackage{siunitx}
\usepackage{float}
\usepackage{xurl}
\usepackage{nicefrac}
\usepackage{array}
\usepackage{multirow}
\usepackage{mathtools}
\usepackage{nccmath}
\usepackage{algorithm}
\usepackage{algpseudocode}
\usepackage{xfrac}

\DeclareSIUnit \dBm {dBm}
\DeclareSIUnit \dB {dB} 
\DeclareSIUnit \dBi {dBi} 
\DeclareSIUnit \Kbps {Kbps}
\DeclareSIUnit \Mbps {Mbps}
\DeclareSIUnit \Gbps {Gbps}
\DeclareSIUnit \Tbps {Tbps}
\DeclareSIUnit \kBps {kBps}
\DeclareSIUnit \MBps {MBps}
\DeclareSIUnit \GBps {GBps}
\DeclareSIUnit \TBps {TBps}

\DeclarePairedDelimiter{\nint}\lfloor\rceil

\DeclarePairedDelimiter{\ceil}{\lceil}{\rceil}

\newcolumntype{N}{>{\centering\arraybackslash}m{3cm}}
\newcolumntype{K}{>{\centering\arraybackslash}m{0.8cm}}
\newcolumntype{P}[1]{>{\centering\arraybackslash}p{#1}}
\newcolumntype{L}[1]{>{\arraybackslash}p{#1}}
\newcolumntype{M}[1]{>{\centering\arraybackslash}m{#1}}

\algnewcommand\algorithmicinit{\textbf{Initialise:}}
\algnewcommand\Init{\item[\algorithmicinit]}

\algnewcommand\algorithmicinput{\textbf{Placement:}}
\algnewcommand\Level{\item[\algorithmicinput]}

\algnewcommand\algorithmicinputttt{\textbf{Output:}}
\algnewcommand\Output{\item[\algorithmicinputttt]}

\def\BibTeX{{\rm B\kern-.05em{\sc i\kern-.025em b}\kern-.08em
    T\kern-.1667em\lower.7ex\hbox{E}\kern-.125emX}}

\begin{document}
\bstctlcite{IEEEexample:BSTcontrol}
\title{Connecting the Unconnected: A DT Case Study of Nomadic Nodes Deployment in Nepal}

\author{
    \IEEEauthorblockN{Ioannis Mavromatis\IEEEauthorrefmark{1}, 
    Angeliki Katsenou\IEEEauthorrefmark{2}, 
    Klodian Bardhi\IEEEauthorrefmark{2}, 
    Evangelos Xenos\IEEEauthorrefmark{2}, 
    and Dimitra Simeonidou\IEEEauthorrefmark{2}}
    \IEEEauthorblockA{
    \IEEEauthorrefmark{1}Digital Catapult, London, U.K.,\\
    \IEEEauthorrefmark{2}University of Bristol, Faculty of Engineering, Bristol, U.K.\\
    Email: ioannis.mavromatis@digicatapult.org.uk, angeliki.katsenou@bristol.ac.uk}
}


\maketitle

\begin{abstract}
This paper addresses the challenge of robust cellular connectivity in dense, underdeveloped urban environments, specifically focusing on Kathmandu, Nepal. As cities grow, existing cellular infrastructure struggles to meet the demand for reliable, high-throughput, and low-latency communication services. The lack of investment in new technologies and the intricacies of the cities' landscape pose even more difficulties for robust connectivity. This work addresses the above challenges in a cost-effective and flexible way. We investigate the deployment of LTE Nomadic Nodes (NNs) at scale in order to enhance network capacity and coverage. Utilising a Digital Twin (DT), we simulate and optimise NN placement, considering Kathmandu's physical and environmental characteristics. Our approach leverages the DRIVE DT framework, which enables the systemic evaluation of various network configurations and user mobility scenarios. The results demonstrate that NNs significantly improve signal strength and expected user datarates, presenting a viable solution for enhancing urban cellular connectivity.
\end{abstract}

\begin{IEEEkeywords}
Digital Twins, Optimal Positioning, LTE, Nomadic Nodes, Dense Urban Environment.
\end{IEEEkeywords}

\section{Introduction}

As urbanisation proliferates, the density of urban environments is unprecedented. Estimates show that by 2050, two-thirds of the world's population will inhabit urban areas~\cite{un17sdg}. Cellular connectivity is crucial for densely populated cities, enhancing urban living through advancements in transportation, healthcare, public safety, and environmental monitoring~\cite{pelton:2019,yaacoub:2020}. Connectivity is also integral for day-to-day activities such as e-banking, staying connected with family and friends, navigation, and infotainment. All use cases require increased throughput, reliable connectivity, and low latency~\cite{5GUseCases}. 

Although the penetration of the public digital infrastructure is increasing in developing countries~\cite{MeasuringDigitalDevelopment_ITU2023,subramanian2006rethinking}, the growing demand for new applications and the increasing number of users outperform the capacity of the infrastructure. The lack of investment in installing additional Base Stations (BSs) worsens the connectivity problem~\cite{Ericsson2023,GSMA2022}.
Addressing these challenges is crucial~\cite{un17sdg}, as a diverse array of critical functions and services that form the socio-economic foundation of contemporary cities depend on seamless connectivity~\cite{dang2020should}.



Inspired by the above, this work will focus on the cellular connectivity problems of Kathmandu, Nepal. We aim to describe a solution that will be cost-effective, easily applicable, and viable in densely populated areas with intricate infrastructure, such as Nepal. We address the problem by investigating the feasibility of creating heterogeneous cellular networks by employing Nomadic Nodes (NNs)~\cite{procBulakci}. NNs are usually self-contained, mobile, and flexible in their deployment and are frequently used to extend existing infrastructure and enhance connectivity across large geographical areas~\cite{procBulakci}.


As in almost all urban areas of Asia-Pacific, Nepal mainly depends on 4G cellular connectivity~\cite{MeasuringDigitalDevelopment_ITU2023, Ericsson2023}. Therefore, we focus on exploring the 4G cellular connectivity in a dense urban scenario in the heart of Kathmandu, Nepal. This scenario needs to be investigated multimodally, considering all the intricacies of dense network deployment, such as user density and mobility, the optimal placement of the NN in conjunction with the existing BSs, the signal attenuation from physical impediments and tall buildings, and more. We employ Digital Twin (DT)-ing, where we model various physical objects in the digital world, emulate different interactions, and run various \emph{what-if} scenarios at a large scale.

Our DT of choice is the Digital Twin for self-dRiving Intelligent VEhicles (DRIVE)~\cite{driveSimulator, DriveGithub}, a framework for large-scale communication scenarios, allowing traffic participants and the communication infrastructure to interact through various communication planes. DRIVE is a ``Digital Network Oracle'' that maintains snapshots of the states of the ``virtual world''. These snapshots can be leveraged for various optimisations or to validate performance changes across large-scale scenarios without the need for resource-expensive simulations~\cite{driveSimulator}. As the current version of DRIVE did not meet the requirements of the investigated scenario, the framework was extended, as discussed in the rest of the paper. 

The NN placement is based on a real map from OpenStreetMaps~\cite{OpenStreetMaps} while taking heed of the existing BS placement extracted from CellMapper~\cite{CellMapper}. Kathmandu's physical landscape (buildings, parks, etc.) is considered for both the optimal placement of NNs and the realistic channel propagation characteristics in such a scenario. Moreover, our dense data traffic conditions are based on spatio-temporal user mobility models to recreate a realistic scenario. DRIVE brings all models and interactions within the same framework, enabling systemic evaluation across the entire city's plane. Our evaluation explores the parameter space, and the results indicate improved network connectivity in terms of the signal strength or the users' perceived datarates.

The rest of the paper is structured as follows. Sec.~\ref{sec:RelatedWork} overviews the state-of-the-art. Our system model is presented in Sec.~\ref{sec:system_model} explaining various extensions to DRIVE. The NN placement and the way we calculate the overall datarate are described in Sect.~\ref{sec:BSPlacement}. Sec.~\ref{sec:Results} summarises our results and findings. Finally, the manuscript concludes in Sec.~\ref{sec:Conclusions}.

\section{Related Work}
\label{sec:RelatedWork}

Traditional macrocell-based mobile communication networks struggle to keep up with the increasing demands of dense urban environments. The densification and deployment of ultra-dense cells aims to enhance the performance of the system and optimise various network key performance indicators (KPIs)~\cite{sym15010002}. Many works tackle that problem with higher-frequency communication planes (e.g., mmWaves)~\cite{mmwaveInfrastructureITS,Asad_SurveyMobility}; however, as seen, such methods require hundreds or thousands of BSs for highly performant deployments. 

However, the cost of such a large-scale fixed telecommunications network is rather prohibitive, particularly in underdeveloped communities where investments are limited~\cite{Ericsson2023,GSMA2022}. Even cost-effective~\cite{costOptimalMmWaves} implementations cannot decrease the cost by a lot. In contrast, the network can be densified using relatively cheap mobile and nomadic small cells to increase coverage and capacity~\cite{MOLLAHASANI2020107271}. NNs exploit their ability to move and establish connections even in previously uncovered zones and dynamically control traffic loads, bridging connectivity gaps in heavily populated areas~\cite{procBulakci}. Moreover, their powerful, on-board processing enables them to perform backhauling operations~\cite{6214336}, still allowing easy relocation, ``plug-and-play'' operation or even functionality while mounted on moving means, e.g., buses, allowing dynamic network formations~\cite{8108598}.

Investigating large-scale cellular scenarios such as the above is enabled with the help of DTs~\cite{dt6G}. DTs are used to investigate inter-cell interference~\cite{Liu2019}, optimise the BS placement~\cite{book_Zhang_2024} and more. However, while many DT approaches are found in the literature, we identified a gap considering the triplet of: 1) realistic propagation models, 2) semantics of the city's infrastructure, and 3) realistic user mobility scenario. Based on the above, our investigation aims to provide a comprehensive investigation of how NNs can enhance connectivity issues in dense urban scenarios such as Kathmandu, Nepal, in a cost-effective and efficient way while considering all the intricacies found in the real world. 


\section{System Model}
\label{sec:system_model}

We consider an urban city map $\mathcal{M}$ with dimensions $\left[ \mathcal{M}_{x}, \mathcal{M}_{y} \right]$, measured in meters. Let $\mathcal{C} \triangleq \left \{1,\ldots,C \right \}$ denote the candidate BS positions, with all being within the boundaries of $\mathcal{M}$. For all the above positions, we denote $\mathcal{C_\mathrm{LTE}} \triangleq \left \{1,\ldots,C_\mathrm{LTE} \right \}$ all the fixed LTE BS positions in our city and $\mathcal{C_\mathrm{NN}} \triangleq \left \{1,\ldots,C_\mathrm{NN} \right \}$ the positions a NN could be placed. We have $\mathcal{C_\mathrm{LTE}}\subseteq\mathcal{C}$, $\mathcal{C_\mathrm{NN}}\subseteq\mathcal{C}$, $\mathcal{C_\mathrm{LTE}}\cup\mathcal{C_\mathrm{NN}}=\mathcal{C}$, and $\mathcal{C_\mathrm{LTE}}\cap\mathcal{C_\mathrm{NN}}=\emptyset$ hold.

In our system model, we assume that all BSs are mounted at the top of a building. Users are either on the road or inside buildings. This section describes how we model users' mobility patterns and their distribution throughout the day (for both outdoor and indoor users). Moreover, we present various models and assumptions introduced in our use case, implemented in DRIVE.

\subsection{Identifying Potential BS Locations}\label{sub:potentialLocations}
\begin{figure}
    \centering
    \includegraphics[width=\columnwidth]{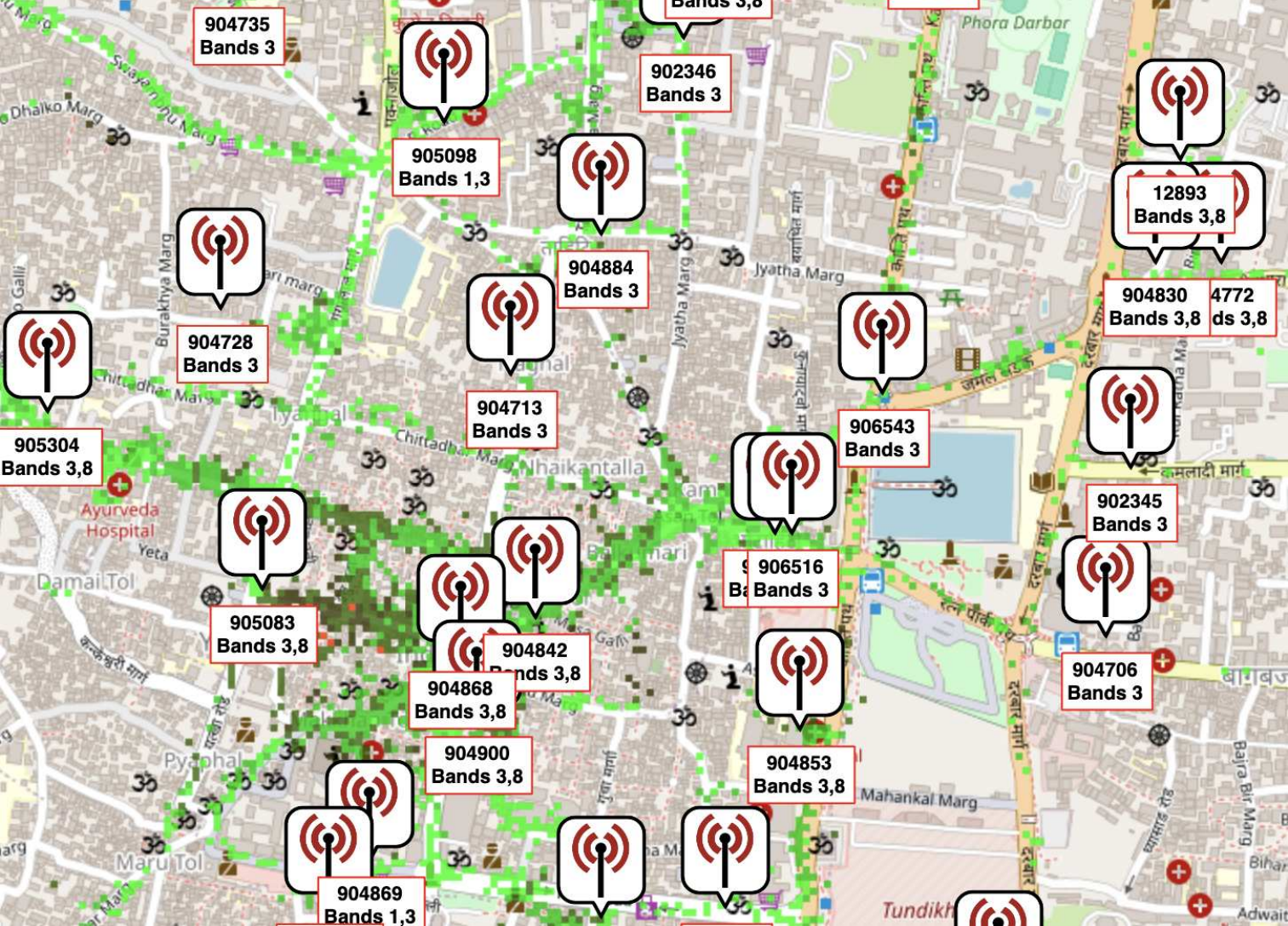}
    \caption{An example of a cropped area from CellMapper and existing BSs.}
    \vspace{-5mm}
    \label{fig:cellmapper}
\end{figure}

We consider two BS deployments, one for LTE and one for NNs. Even though both LTE BSs and the introduced NNs are modelled upon the same LTE release, for simplicity, we refer to them as LTE and NN for the rest of the paper. The existing LTE infrastructure $\mathcal{C_\mathrm{LTE}}$ is extracted using CellMapper~\cite{CellMapper}. CellMapper is community-driven and designed to map cellular network coverage by identifying cell tower locations globally. Users who volunteer to use its mobile application passively report data, such as frequency band and bandwidth. CellMapper aggregates these data and presents the positions, real-time information and road coverage of existing BSs across different service providers. Crowdsourced data are invaluable for network optimisations and identification of coverage gaps, such as in our work. An example of the CellMapper interface is shown in Fig.~\ref{fig:cellmapper}. Nepal, is served primarily by two major telecommunication providers, NCell and NTCell, who occupy more than $94\%$ of market share~\cite{Dahal_2019}. Therefore, our investigation is based on extracted BS locations from those providers. The NN placement is described in Sec.~\ref{sec:BSPlacement}.

\begin{figure}[t]     
    \centering
    \includegraphics[width=1\columnwidth]{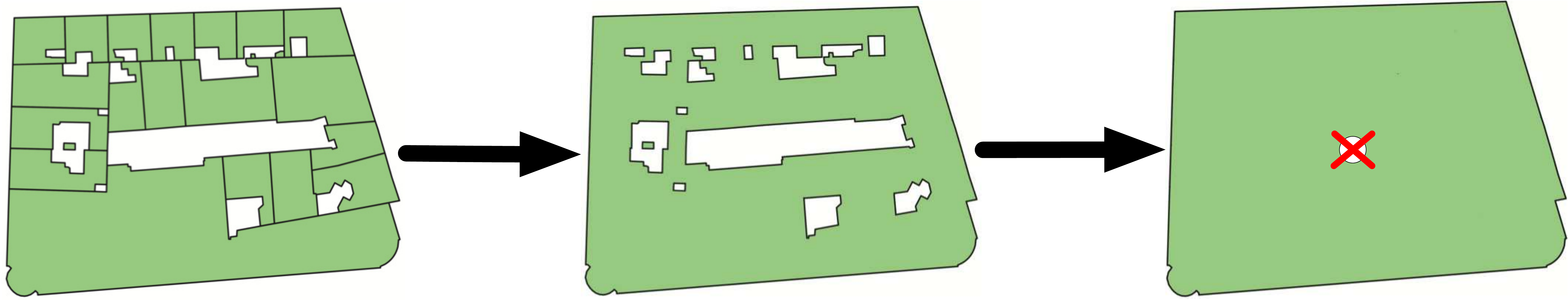}
    \caption{Example of the polygon union operation for a given city block and the calculated geometric in-centre.}
    \vspace{-5mm}
    \label{fig:unionPolygonOper}
\end{figure}

All BSs are considered to be mounted on the rooftop of a building at its in-centre (Fig.~\ref{fig:unionPolygonOper}). To identify building positions, we use OpenStreetMap~\cite{OpenStreetMaps}. The map exported is parsed, and the buildings, roads, street furniture, parks, and more are manipulated using Computational Geometry tools. The buildings particularly are represented as 3D Simple Polygons (SPs).
An SP, seen in the 2D space, is a flat-shaped object consisting of straight, non-intersecting line segments that, when joined pair-wise, form a closed path. Under urban scenarios, building blocks usually consist of buildings with adjacent tangent sides or small negligible gaps between them. Such building blocks are concatenated for our experimentation using the polygon union operation~\cite{polygonUnion}. Holes formed from this concatenation (e.g., a courtyard) are later removed, forming a solid object (as in Fig.~\ref{fig:unionPolygonOper}). For the remainder of the paper, an SP in 3D is described as a ``building'' and $\mathcal{B} \triangleq \left \{1,\ldots,B \right \}$ denotes all buildings in our scenario; the 3rd dimension is represented with a random height $h_\mathrm{B}$ given to each $B$.

As the positions extracted from CellMapper are triangulated based on user-reported data, they may not perfectly align with a given building in our simulator. Therefore, within DRIVE, we align the parsed positions with the nearest building's in-centre (the polygon's geometric centre). Similarly, all remaining building centres are considered candidate positions for placing an NN. The SPs are also used to determine whether a link is in Line-of-Sight (LOS) or Non-Line-of-Sight (NLOS). The road polygons are similarly manipulated. The metadata accessible from OpenStreetMap provide the information (e.g., lanes, width, etc.) for the roads, and concatenating the polygons, we form concave polygons with holes. Similar polygons are formed for open areas around the city (e.g., parks).

\subsection{LOS and NLOS Tile Tesselation}{\label{subsec:los_nlos}}
The building polygons are utilised to identify LOS and NLOS links. Given $\mathcal{M}$, we model the city by equally spacing $\mathcal{Z} \triangleq \left \{1,\ldots,Z \right \}$ grid points on the map with equal weights. Each point represents a squared tile with the same Received Signal Strength (RSS) throughout its surface. Using a tile-like approach with relatively small tiles can decrease the processing power required without significant loss in accuracy.

When overlaying these tiles on a map, all tiles within road polygons, parks or open-air districts are considered the ``outdoor'' area, and we denote them as $\mathcal{Z_\mathrm{out}} \triangleq \left \{1,\ldots,Z_\mathrm{out} \right \}$. Tiles within building polygons are considered ``indoors'' and are indicated as $\mathcal{Z_\mathrm{in}} \triangleq \left \{1,\ldots,Z_\mathrm{in} \right \}$. We have $\mathcal{Z_\mathrm{out}}\subset\mathcal{Z}$, $\mathcal{Z_\mathrm{in}}\subset\mathcal{Z}$, $\mathcal{Z_\mathrm{out}}\cup\mathcal{Z_\mathrm{in}}\subseteq\mathcal{Z}$, and $\mathcal{Z_\mathrm{out}}\cap\mathcal{Z_\mathrm{in}}=\emptyset$ hold. For $\mathcal{Z_\mathrm{out}}$, we can calculate whether they are in LOS or NLOS, taking a straight line between the tile's in-centre and a given BS position. A link is considered NLOS if at least one intersection with a building polygon exists. The polygon on which a BS is mounted is always excluded from this calculation. Finally, all ``indoor'' users are always considered in NLOS. For simplicity, all ``indoor'' users are considered to be at the building in-centre, and all users are assigned the same height $h_\mathrm{u}$.

\subsection{Link Budget Analysis}

To calculate our link budget, under LOS and NLOS conditions, the \emph{RSS} is given as follows:
\begin{equation}
P_\mathrm{rx} = P_\mathrm{tx} + G_\mathrm{tx} + G_\mathrm{rx} - PL
\end{equation}
where $P_\mathrm{rx}$ and $P_\mathrm{tx}$ are the received and transmitted power (in \si{dBm}), respectively, and $G_\mathrm{rx}$ and $G_\mathrm{tx}$ are the RX and TX antenna gains (in \si{dB}). Finally, $PL$ is the \emph{path-loss component} (in \si{dBm}). For our path loss propagation, we use the LOS and NLOS models for urban macrocell environments described in 3GPP LTE Release 17~\cite{3GPP38901}, defined as:
\begin{align}
\begin{split}
PL_\mathrm{LOS} = & \, 28 + 40 \, \log_{10}d + 20 \,  \log_{10}f_\mathrm{c} \\
                & -  9 \, \log_{10}(d_\mathrm{BP}^2 + (h_\mathrm{B} - h_\mathrm{u})^2) + SF
\end{split}
\\
\begin{split}
PL_\mathrm{NLOS} = & \, 13.54 + 39.08 \, \log_{10}d + 20 \,  \log_{10}f_\mathrm{c} \\
                & -  0.6 \, (h_\mathrm{u} - 1.5)  + SF
\end{split}
\end{align}
where $d$ is the distance separation (in \si{\meter} in a 3D space) between a BS and a user, $f_\mathrm{c}$ is the carrier frequency (in \si{\giga\hertz}), and  $SF$ is the random shadowing effect  (in \si{\dB}) following a log-Normal distribution $SF\sim\log\mathcal{N} (0,\sigma^2)$, with $\sigma_\mathrm{LOS}$ being $4$ and $\sigma_\mathrm{NLOS}$ equal to $7.8$~\cite{3GPP38901}. The parameter $d_\mathrm{BP}$ represents the breakpoint distance and is $d_\mathrm{BP} = h_\mathrm{B} \times h_\mathrm{u} \times \nicefrac{f_\mathrm{c}}{c}$ with $c$ the propagation velocity in free space and equal to  $c = 3\times10^8 \, \si{\meter\per\second}$.

\begin{table}[t]
\renewcommand{\arraystretch}{1.2}
\centering
    \caption{MCSs and sensitivity thresholds for SISO.}
    \begin{tabular}{|M{0.8cm}||M{1.4cm}|M{1.1cm}|M{1.7cm}|M{1.4cm}|}
    \hline
    \textbf{RAT} & \textbf{Modulation} & \textbf{Coding Rate} & \textbf{Datarate} & \textbf{$\mathbf{L_\mathrm{MCS}}$ Threshold} \\ \hline \hline
    \parbox[t]{2mm}{\multirow{4}{*}{\rotatebox[origin=c]{90}{LTE}}} & QPSK & 0.4385 & \SI{18.9}{\Mbps} & \SI{-92.2}{\dB} \\ 
    & 16-QAM & 0.6016 & \SI{27.8}{\Mbps} & \SI{-81.2}{\dB} \\ 
    & 64-QAM & 0.8525 & \SI{56.7}{\Mbps} & \SI{-75.2}{\dB} \\ 
    & 256-QAM & 0.9258 & \SI{75.6}{\Mbps} & \SI{-70.2}{\dB} \\ \hline \hline
    \parbox[t]{2mm}{\multirow{4}{*}{\rotatebox[origin=c]{90}{Nomadic}}}  & QPSK & 0.4385 & \SI{25.2}{\Mbps} & \SI{-91}{\dB} \\
    & 16-QAM & 0.6016 & \SI{50.4}{\Mbps} & \SI{-80}{\dB} \\ 
    & 64-QAM & 0.8525 & \SI{75.6}{\Mbps} & \SI{-74}{\dB} \\
    & 256-QAM & 0.9258 & \SI{100.8}{\Mbps} & \SI{-69}{\dB} \\ 
    \hline
	\end{tabular}
\label{tab:mcs_sinr}
\end{table}

\begin{figure*}[t]
    \centering
    \begin{minipage}[t]{0.47\textwidth}
        \begin{subfigure}[t]{\textwidth}
            \centering
            \raisebox{0.18\height}{\includegraphics[width=0.98\textwidth]{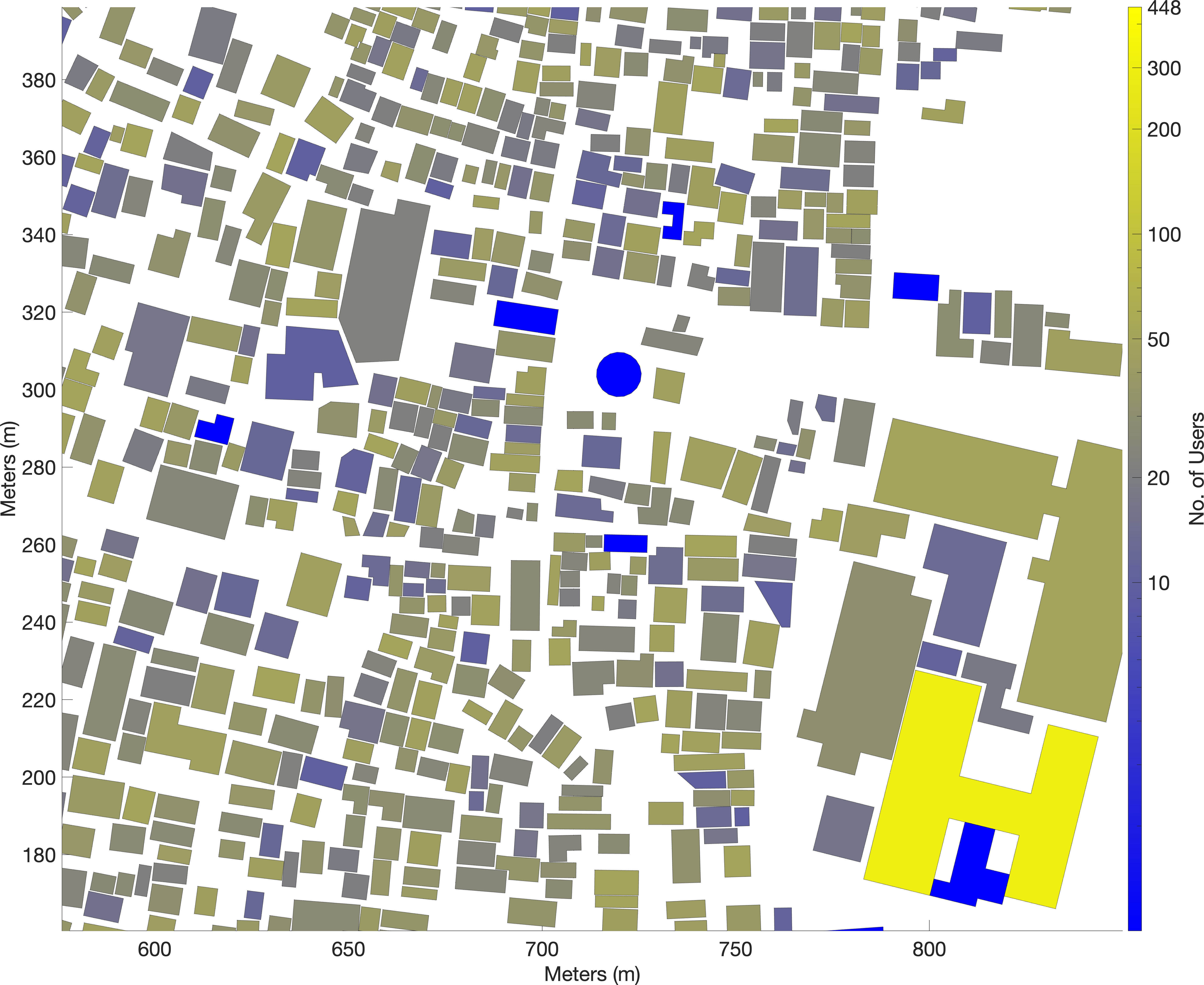}}
            \vspace{-12mm}
            \caption{An example user distribution per building (dark blue indicates the lowest number of users, while bright yellow is the highest).}
            \label{fig:building_distribution}
        \end{subfigure}
    \end{minipage}%
    \hspace{0.05\textwidth}%
    \begin{minipage}[b]{0.48\textwidth}
        \centering
        \begin{subfigure}[b]{\textwidth}
            \centering
            \includegraphics[width=\textwidth]{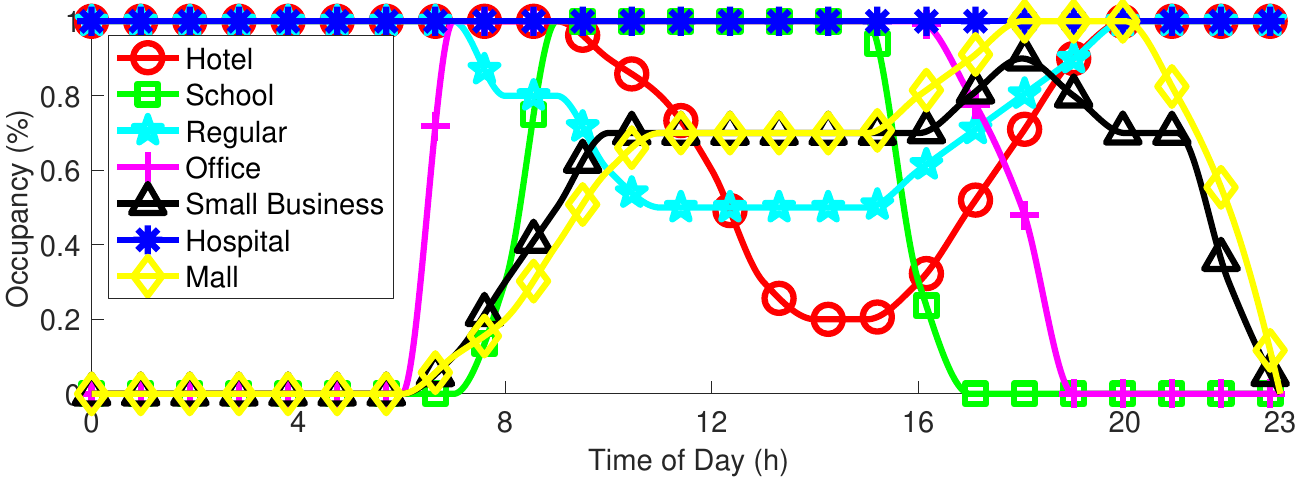}
            \caption{Presence of users within each building throughout the day.}
            
            \label{fig:occupancy}
        \end{subfigure}
        \vspace{0.05\textwidth}
        \begin{subfigure}[t]{\textwidth}
            \centering
            \includegraphics[width=\textwidth]{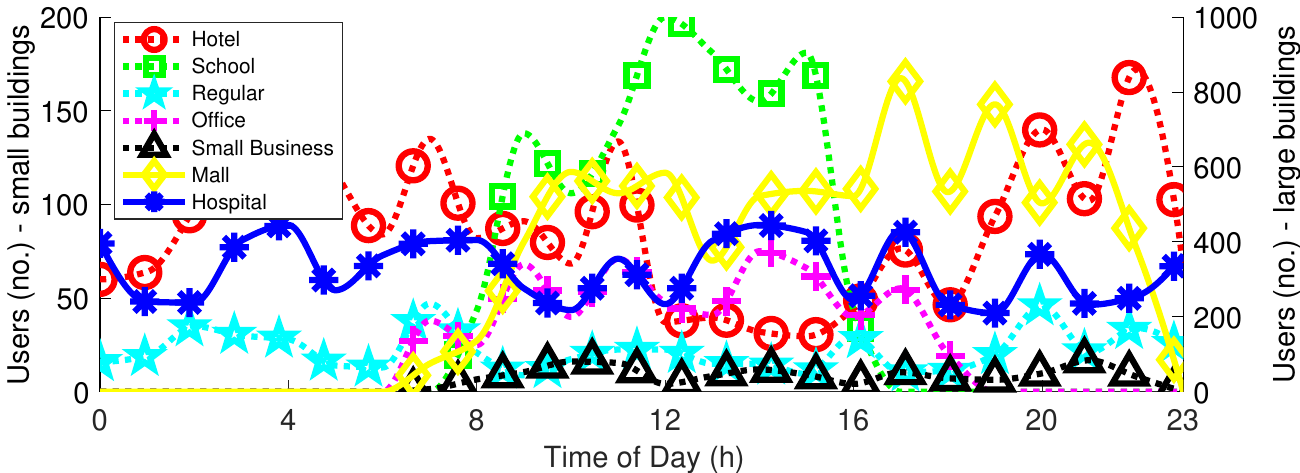}
            \caption{Example user distribution across an entire day for different building types. 
            }
            \label{fig:no_of_users_buildings}
        \end{subfigure}
    \end{minipage}
    \vspace{-6mm}
    \caption{Examples of the user distribution for all building types throughout a day.}
    \vspace{-5mm}
    \label{fig:buildings_traffic}
\end{figure*}

\subsection{Datarate and Link Adaptation}\label{subsec:datarate_adaptation}
LTE performs a link adaptation and selects the appropriate Modulation and Coding Scheme (MCS) based on a Channel Quality Indicator (CQI), delay and throughput requirements, and historical data. For our model, we perform a simple link adaptation based on the link quality described above and the sensitivity level $L_\mathrm{MCS}$ of each MCS and bandwidth. The sensitivity levels are found from the ETSI TS 136 101 standard~\cite{3gpp-ts-36-101-v17-6-0}. We map all sensitivity levels with an MCS (one-to-one), with the sensitivity level being the lowest link quality (lowest RSS) that the channel can support the given MCS. Later, we calculate the theoretical PHY datarate for all MCSs and a given channel bandwidth $BW_\mathrm{c}$ (measured in \si{\hertz}). 

Comparing the $P_\mathrm{rx}$ with the sensitivity levels, we identify the MCS to be used and calculate the theoretical PHY datarate for a given tile as follows. Each LTE Physical Resource Block (PRB) has 12 subcarriers (each subcarrier is \SI{15}{\kilo\hertz}) in the frequency domain and \SI{0.5}{\milli\second} (7 symbols) in the time domain~\cite{etsi_ts_136211}. Thus, the total number of symbols for one PRB is $s_\mathrm{PRB} = 12 \times 7$. We also know the available resource blocks $PRB_\mathrm{a}$ for different $BW_\mathrm{c}$~\cite{etsi_ts_136211}. Based on that, we calculate the symbols for the entire channel, i.e. $s_\mathrm{BW_\mathrm{c}} = PRB_\mathrm{a} \times s_\mathrm{PRB}$. From the 3GPP's standard~\cite{etsi_ts_136211}, each subframe is \SI{1}{\milli\second}, equating to $2$ time slots. Thus, the number of symbols per \SI{1}{\milli\second} is $2 \times s_\mathrm{BW_\mathrm{c}}$. We know the bits per symbol $b_\mathrm{s}$ for all MCSs. Based on that we calculate the datarate $D_\mathrm{MCS}$ for each MCS with $D_\mathrm{MCS} = 2 \times s_\mathrm{BW_\mathrm{c}} \times b_\mathrm{s} \times 1000 \times (1 - O)$, measured in \si{\bit\per\second}. $O$ is an approximated overhead (in $\%$) used for controlling and signalling. Table~\ref{tab:mcs_sinr} shows an overview of all MCSs for a SISO channel, their $D_\mathrm{MCS}$ and $L_\mathrm{MCS}$. For MIMO, the datarate becomes $L \times D_\mathrm{MCS}$, where $L$ is the number of antenna elements in the TX and RX sides. 

This is the theoretical maximum datarate achieved if only a single user existed in a scenario. 4G LTE cells are split in sectors depending on the number of carriers the hardware supports. For $1$ carrier, the cell is discretised into $3$-sector sites with \ang{120} coverage per sector, while for $2$ carriers, the beamwidth $\theta$ becomes \ang{60} and the sectors are increased to $6$. Within such a configuration, LTE can accommodate $D_\mathrm{MCS}$ across each sector, multiplying the maximum link capacity and the number of total concurrent users per cell. CellMapper shows that BSs at Kathmandu, Nepal, primarily support $3$ sectors, whereas the modelled NN supports $6$ sectors.

\subsection{User and Traffic Mobility Scenarios}

Modelling and simulating a cellular network typically assumes a user or a traffic distribution across BS cells. Works in the literature tend to uniformly scatter users across an area. However, traffic patterns are highly non-uniform across different cells~\cite{spatialModelling}. Moreover, the traffic distribution across cells does not indicate the real spatial traffic distribution, as cell coverage can be highly uneven. Therefore, the traffic density per unit area is a more realistic representation of traffic demand in a cellular network. The temporal aspect of cellular traffic is also paramount, as described in~\cite{temporalModelling}. For example, during daytime, users commute from residential areas to office buildings, visit recreational areas during lunch breaks, go shopping after work, or return to their residencies at night. This user behaviour pattern results, for example, in high daytime traffic volume and low nighttime traffic volume in office districts or the opposite in residential areas. The traffic demand, therefore, is by nature spatio-temporal. Thus, we define two traffic distributions, one for users within buildings and another for users on the road. We model our cellular traffic as a function of the number of users within a unit of area as $u(t)_{\mathrm{Z}}$, i.e., a tile $Z$ as described in Sec.~\ref{subsec:los_nlos}, or within a building $u(t)_{\mathrm{B}}$, with $t$ being the time of day (in \si{\hour}). 

\begin{figure*}[ht]
    \centering
    \begin{subfigure}[b]{0.24\textwidth}
        \centering
        \includegraphics[width=\textwidth]{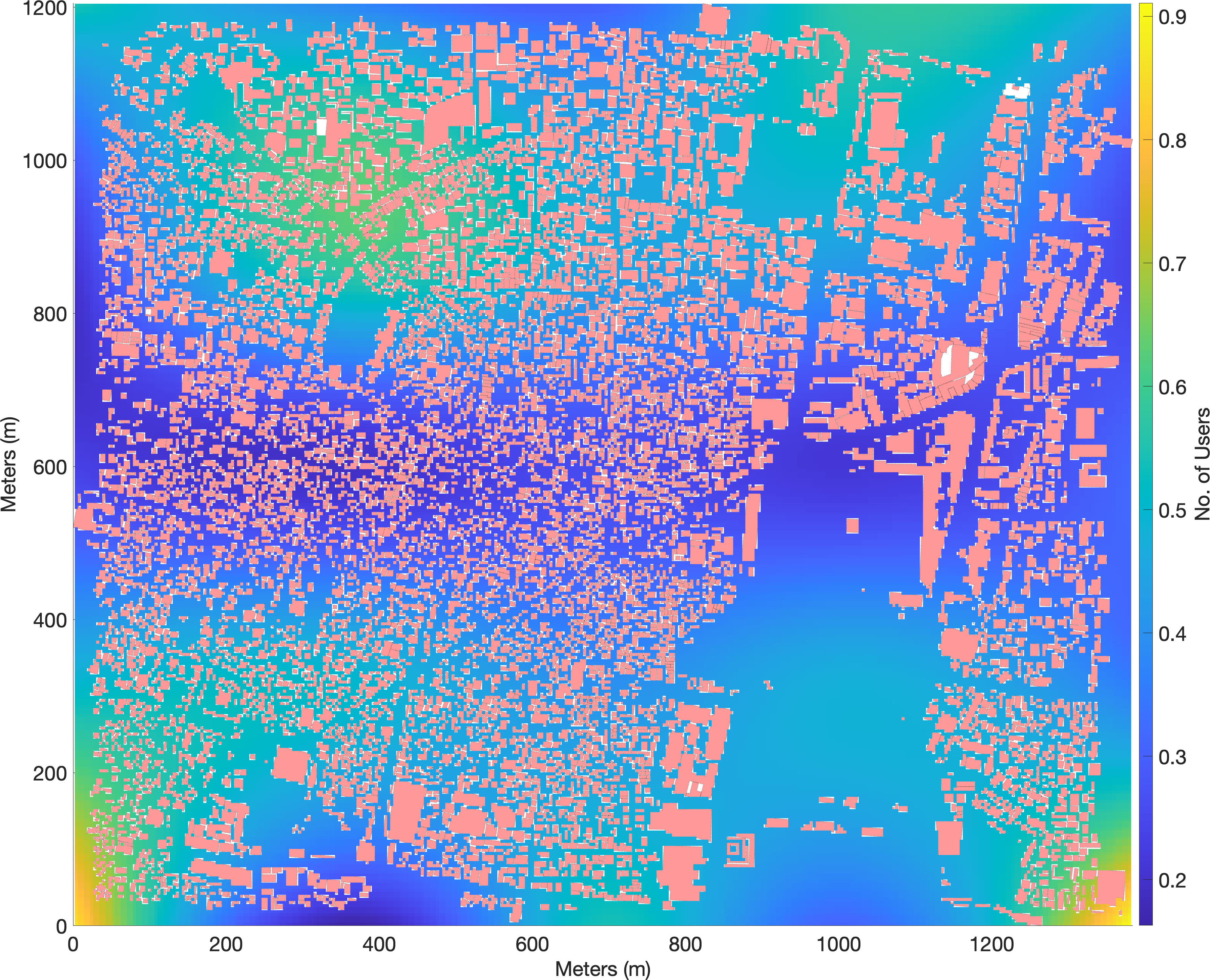}
        \caption{2am}
        \label{fig:2am}
    \end{subfigure}
    \hfill
    \begin{subfigure}[b]{0.24\textwidth}
        \centering
        \includegraphics[width=\textwidth]{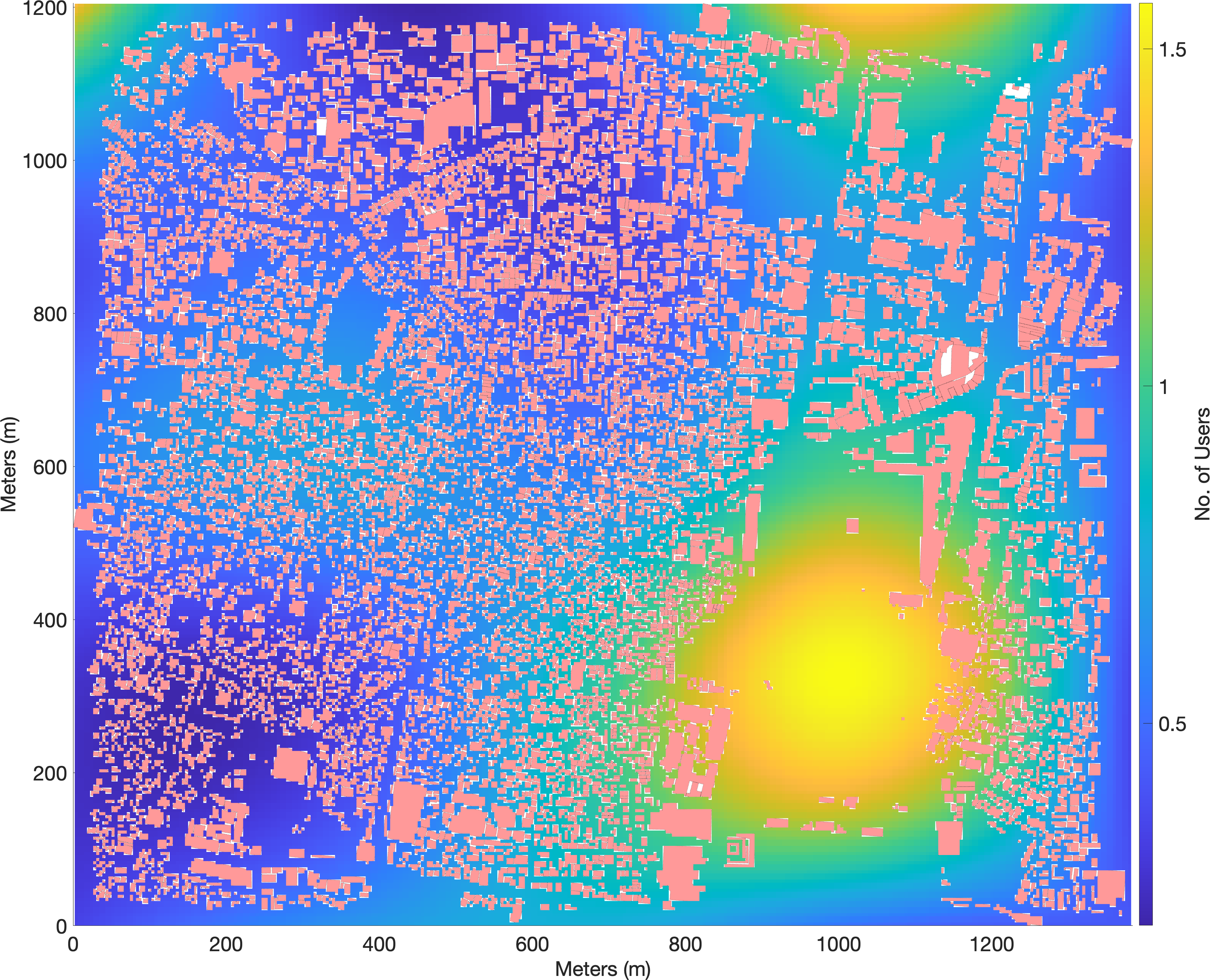}
        \caption{8am}
        \label{fig:8am}
    \end{subfigure}
    \hfill
    \begin{subfigure}[b]{0.24\textwidth}
        \centering
        \includegraphics[width=\textwidth]{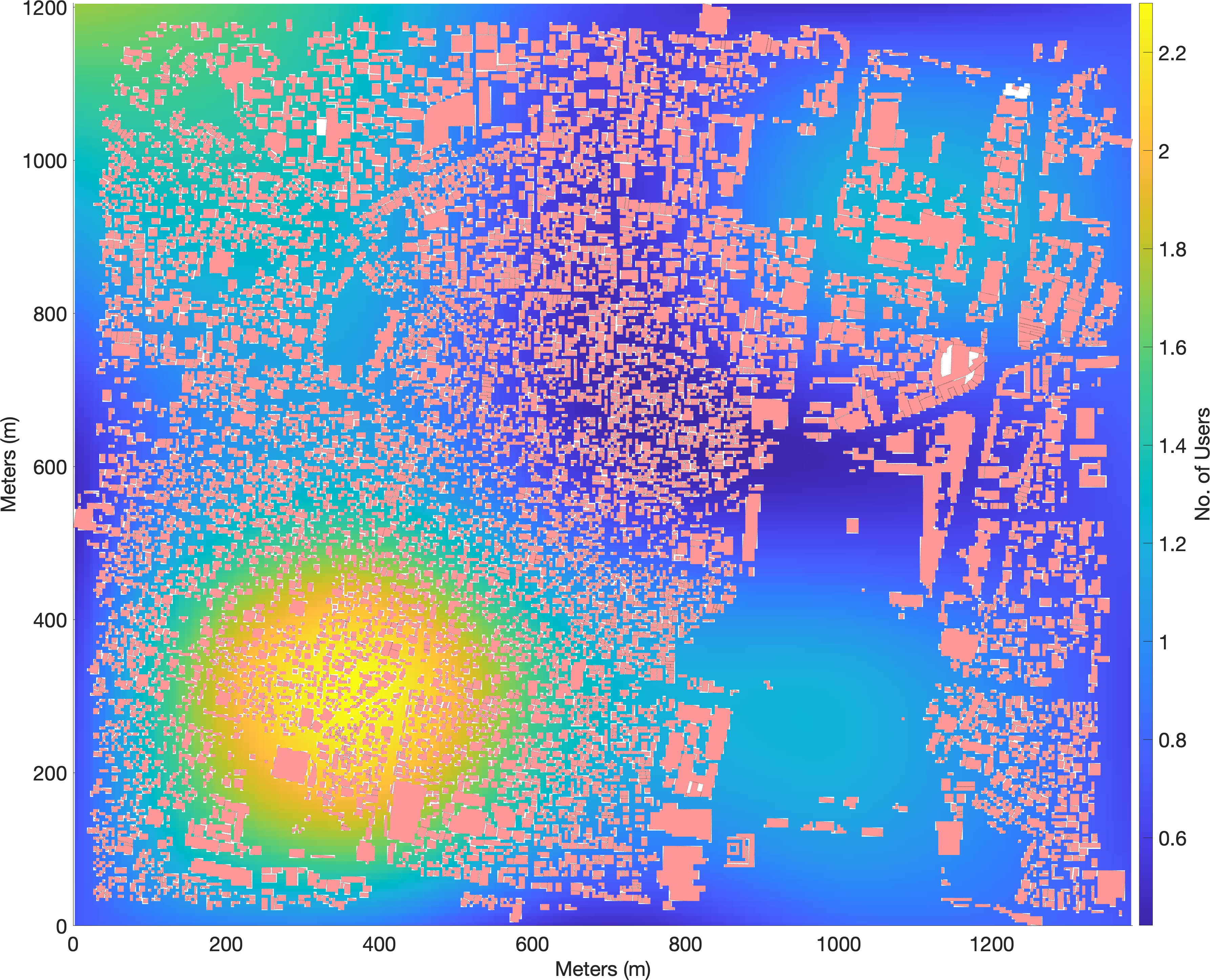}
        \caption{4pm}
        \label{fig:4pm}
    \end{subfigure}
    \hfill
    \begin{subfigure}[b]{0.24\textwidth}
        \centering
        \includegraphics[width=\textwidth]{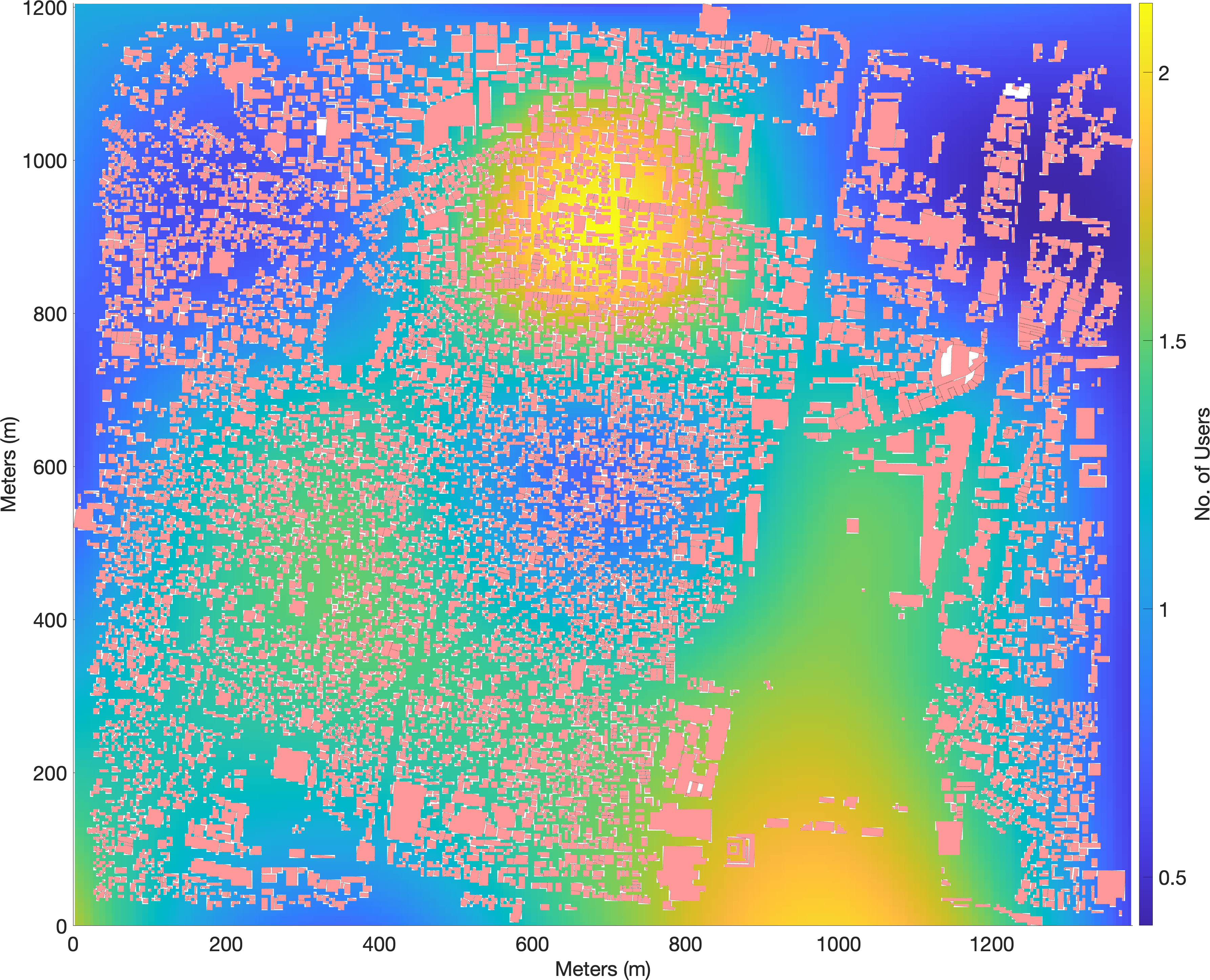}
        \caption{10pm}
        \label{fig:10pm}
    \end{subfigure}
    \caption{An example of ``road'' users density throughout the day. The colourbar shows the user density per $4\times4\si{\meter}^2$ areas.}
    \label{fig:traffic_roads}
\end{figure*}

\subsubsection{Users in Buildings}
OpenStreetMap metadata contain information on the building types. A building type can be $\kappa \in \left\{\right.\mathrm{Hotel}$, $\mathrm{School}$,  $\mathrm{Residential}$, $\mathrm{Office}$, $\mathrm{Small \, Business}$, $\mathrm{Hospital}$, $\mathrm{Mall} \left.\right\}$. Hospitals and malls are considered large buildings, while the rest are small. Each building type is assigned a range of users, and a random value $u_\kappa \in \mathbb{N}^* \cap [u_1,u_2]$ is picked at a given time $t$, defining the maximum potential occupancy for this building. The chosen ranges are: $u_\mathrm{Hotel} \in [50, 200]$, $u_\mathrm{School} \in [100, 200]$, $u_\mathrm{Residential} \in [10, 50]$, $u_\mathrm{Office} \in [25, 75]$, $u_\mathrm{Small \, Business} \in [5, 30]$, $u_\mathrm{Hospital} \in [200, 450]$, $u_\mathrm{Mall} \in [500, 1000]$. These values are based on various reports from neighbourhoods similar to Kathmandu, Nepal and interactions with locals. For each $\kappa$ and $t$, we also define an occupancy rate $\rho(t)_\kappa$, representing the utilisation of a building according to the time of day. As before, the occupancy rates chosen are based on local sources. The indoor users for a building at a given $t$ are then calculated as $u(t)_{\mathrm{B_\kappa}} = \rho(t)_\kappa \times u_\kappa$. Fig.~\ref{fig:buildings_traffic} illustrates instances of the generated user distributions based on the spatio-temporal modelling. In Fig.~\ref{fig:building_distribution}, the varying density per building is illustrated using dark blue for low user density and bright yellow for high. For example, the bright yellow building in the bottom right corner corresponds to a hospital that is assumed to always have high occupancy compared to other buildings. This is also demonstrated in Fig.~\ref{fig:occupancy}, which shows the occupancy rates chosen for our scenario for all building types. For example, buildings types such as offices or schools are expected to have minimum occupancy during nighttime but peak occupancy during working hours, as shown in Fig.~\ref{fig:no_of_users_buildings}

\subsubsection{Users on the Road}
For road users, the traffic modelling in~\cite{temporalModelling} implements the temporal aspect patterns. Mobile users' traffic density has three main frequency components, at $\pi/12$, $\pi/6$ and $\pi/4$, which correspond to the periods of \SI{24}{\hour}, \SI{12}{\hour} and \SI{8}{\hour}, respectively. These show repetitive behaviours of users in periods of one day, half a day and working hours. Thus, a 3rd-order sinusoid waveform was proposed that fits real traffic patterns in dense urban scenarios:
\begin{align}\label{eq:traffic_pattern}
\begin{split}
V_t^\prime = & \,  173.29 + 89.83 \, \sin(\frac{\pi}{12}t^\prime + 3.08) \\
      & \, +  52.6\, \sin(\frac{\pi}{6}t^\prime + 2.08) + 16.68 \, \sin(\frac{\pi}{4}t^\prime + 1.13)
\end{split}
\end{align}
with $t^\prime$, being a given time of day (in \si{h}). As our model is based on the users per tile $u(t)_{\mathrm{Z}}$, we define $\max u_\mathrm{Z}$ the maximum number of users that can be found at any given $t$, at any given $Z$. We later normalise equation~\eqref{eq:traffic_pattern} as:
\begin{equation}\label{eq:norm_v}
    \mathrm{norm}V_t^\prime = \nint*{1 + \frac{(V_t^\prime - \min V_t^\prime)}{\max V_t^\prime - \min V_t^\prime} * (\max u_\mathrm{Z} - 1))}
\end{equation}
that gives us a range of $u_{\mathrm{Z}} \in \mathbb{Z} \cap [1,\max u_\mathrm{Z}]$, $\forall Z \in Z_{\text{out}}$. Authors in~\cite{temporalModelling} do not explicitly describe when peak or off-peak times are shown during a given day. A spatio-temporal analysis of cellular traffic in metropolises~\cite{timeAlignment} shows that the lowest traffic density occurs at ``$4$ a.m.'' and their traffic patterns align with Eq.~\eqref{eq:traffic_pattern}. Therefore, for our scenarios, we synchronise $V_t^\prime$ with our timeframe $t$, ensuring that the lowest user density is at ``$4$ a.m.'' and getting our final $V_t$.

The spatial aspect of user density on the roads is inspired by~\cite{spatialModelling}, where the spatial traffic density is modelled for dense urban scenarios. The model is built by a sum of sinusoids that captures the characteristics of log-normally distributed and spatially correlated cellular traffic. A log-normal distribution is usually unimodal with a single ``peak'' (i.e., a global maximum). However, as described in~\cite{jiang2024spatiotemporal}, traffic density in dense urban environments shows several hotspot areas even within maps that are $\leq\SI{1}{\kilo\meter}^2$. Combining the above, we build a model that extends the one from~\cite{spatialModelling} and introduces multiple hotspots within our scenario. More specifically, we define a parameter $N \in \mathbb{N}^*$ as $N = \ceil{(\mathcal{M}_{x} \times \mathcal{M}_{y})/420}$ with $\SI{420}{\meter}^2$ being an indicative area where at least one hotspot can appear within a dense urban environment~\cite{cesario2024detecting}. We calculate:
\begin{equation}
    u(t)_\mathrm{Z} = \frac{\sum_{x=1}^{M_x}\sum_{y=1}^{M_y} F(x,y)}{N} \, V_t
\end{equation}
with $F(x,y)$ being the spatial model defined in~\cite{spatialModelling} and calculated as a Gaussian random field:
\begin{align}
r(x,y) = \frac{2}{\sqrt{L}} \sum_{l=1}^{L} \, \cos(i_l \, x + \phi_l) \, \cos(j_l \, y + \psi_l)
\end{align}
where $x,y$ are the Cartesian coordinates of tile $\mathrm{Z}$, $i_l$ and $j_l$ are uniform random variables between $0$ and the maximum spatial spread $\omega_\mathrm{max}$, which decides the rate of fluctuations of the random field. $L$ is a user defined variable and $\phi_l$ and $\psi_l$ are uniform random variables $\in [0,2\pi]$. Finally, taking the exponential function, we calculate the traffic density as a log-normal distribution:
\begin{equation}
F(x,y) = \exp(\sigma r(x,y) + \mu), \;\; \forall x,y \, \in \, \mathcal{M}_{x}, \mathcal{M}_{y}
\end{equation}
Fig.~\ref{fig:traffic_roads} illustrates an example of the generated user density. As seen throughout the day, multiple peaks are formed across the plane, and the number of users changes following Eqs.~\eqref{eq:traffic_pattern} and~\eqref{eq:norm_v}. The user density in Fig.~\ref{fig:traffic_roads}, is a function of $u(t)_{\mathrm{Z}}$ within every tile $Z$ in the city.


\subsection{Nomadic Node}\label{subsec:nomadic_node}
NNs fulfil the need for cellular connectivity in areas with limited infrastructure that would benefit from a private network, such as ports, arenas and parks. Beyond connectivity, the NN can offer Mobile Edge Computing (MEC) capabilities for computationally intensive services, such as live video streaming. It essentially houses all necessary networking and hardware components, including wireless technologies like WiFi, to create a mobile data centre solution.

\begin{figure}
   \centering
    \includegraphics[width=.78\columnwidth]{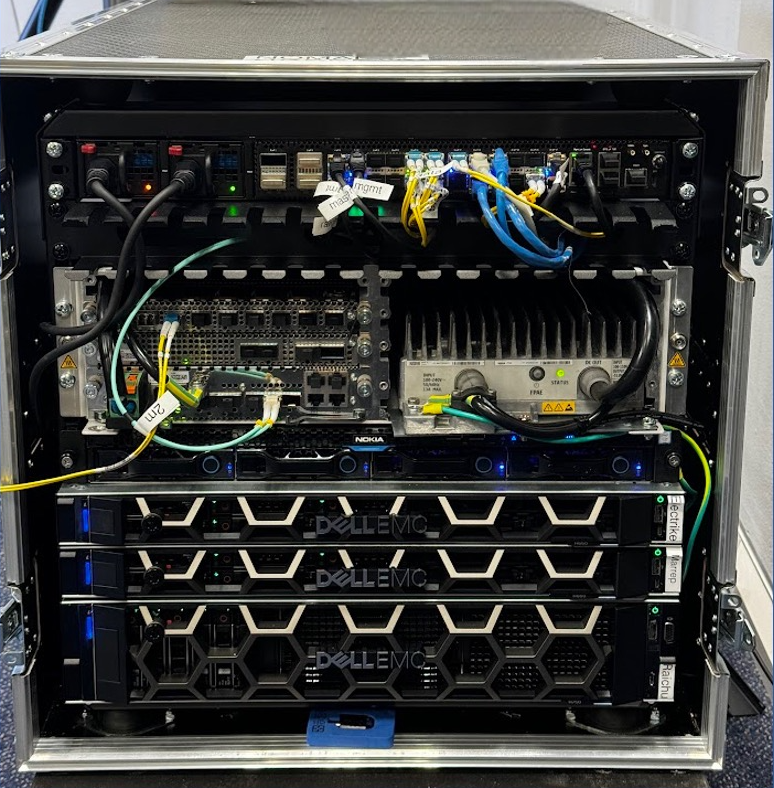}
   \caption{A photo of the hardware setup of the NN.}
   \label{fig:NNode}
\end{figure}

Fig.~\ref{fig:NNode} depicts the hardware setup of an NN. The presented NN comprises servers, switches, routers, 4G/5G Radio Access Network (RAN), and wireless Access Point (AP) technologies, such as WiFi and LiFi. The servers host various technologies confined in virtualised environments, enhancing the usability and overall performance of the infrastructure through a series of AI/ML-powered applications, such as the Multi-Access Technology Real-Time Intelligent Controller (mATRIC), created by the University of Bristol as part of the REASON project~\cite{mATRIC}. The main body of the hardware setup can be placed in a secure outdoor or indoor location. The 4G/5G radio, connected via fibre, can be deployed at a distance (even several kilometres away). The versatility of its deployment has been showcased in past projects such as the 5G-VICTORI~\cite{5g-victori:2021}, where a NN was deployed on a historic boat in Bristol to provide connectivity for users enjoying a river cruise. 

The NN can leverage existing broadband connections (e.g., from offices, houses, public spaces, or satellite links) and, in an almost ``plug-and-play'' fashion, spin up a private 4G/5G network. For simplicity, handover mechanisms are out of the scope of this work, and we assume that it will happen seamlessly between NNs and LTE BSs. Finally, all the characteristics and specs of this NN are modelled within DRIVE.



\section{BS Placement and Perceived Datarate}
\label{sec:BSPlacement}
Telecommunication providers follow various strategies for dense urban BS placements depending on waveform, timing, or power observations. For example, BS positions that maximise coverage may be preferred, hotspots or points of interest that generate increased traffic may be prioritised, and device characteristics and wall penetration are considered. In dense urban environments such as Kathmandu, Nepal, the infrastructure may be outdated, and the adoption of new technologies may be prohibitive because of the high cost. Also, scaling the deployment may be difficult because of buildings' structural integrity, limited rooftop space, or power supply limitations. 

\begin{figure}
       \includegraphics[trim={55 132 54 130}, clip, width=\columnwidth]{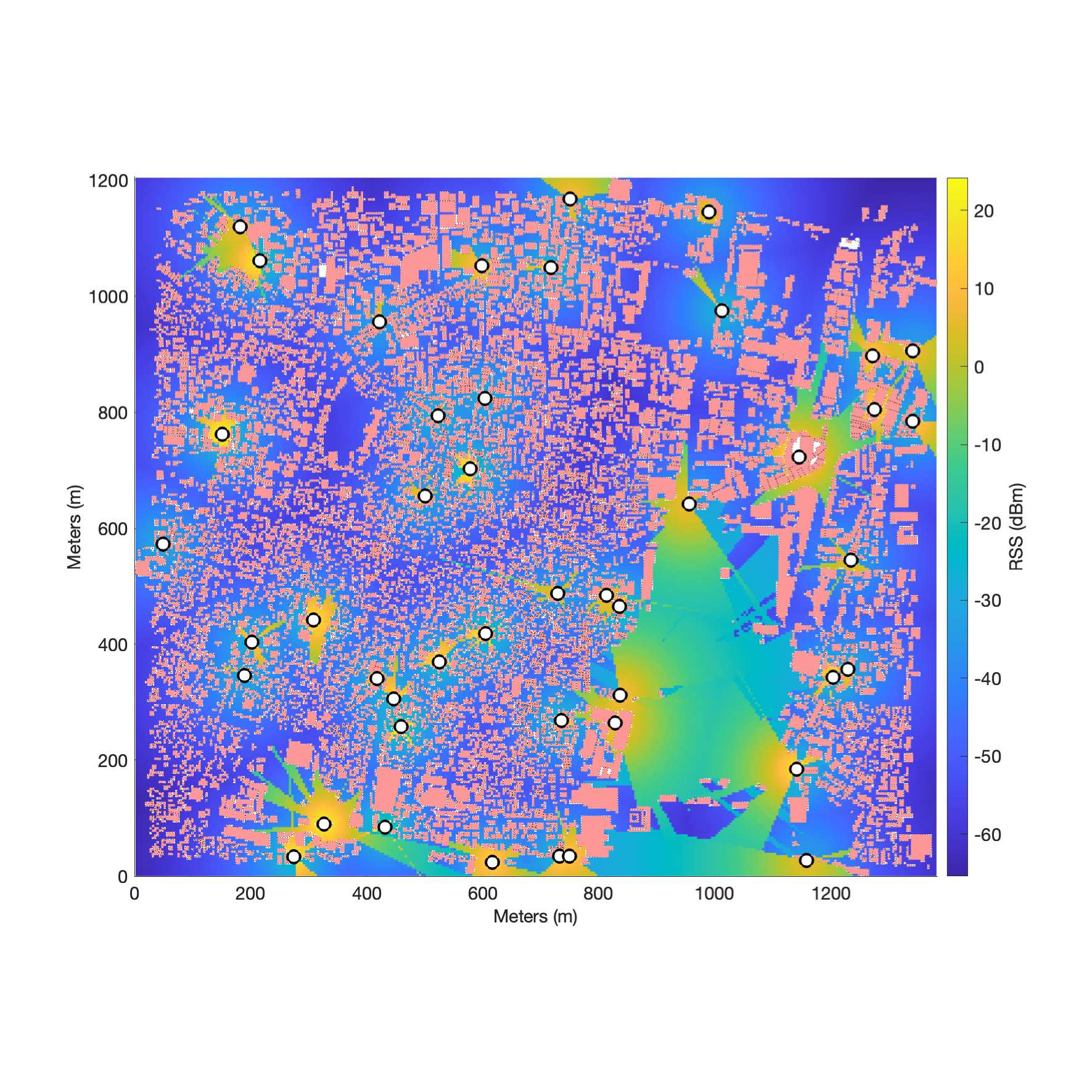}
    \caption{RSSI heatmap baseline (produced with existing BSs).}
    \label{fig:rssi_existing}
\end{figure}

NNs like the one introduced in Sec.~\ref{subsec:nomadic_node} are designed for areas that lack wireless technologies and provide a compact, cost-effective solution. Investigating the RSS (Fig.~\ref{fig:rssi_existing}) in the area of interest, we identify that the coverage achieved by the existing infrastructure is fair on average. This result aligns with the coverage observed on CellMapper (Fig.~\ref{fig:cellmapper}) and reflects the user experience from locals. From discussions, it was conveyed that the link capacity is of the primary concern, with the datarate rapidly decreasing, particularly during day time. In wireless communications, users near the edge of a coverage area have a higher probability of errors and retransmissions, consuming additional resources and leading to increased delays and lower throughput. The increased time-frequency resource allocation (e.g., time slots, subcarriers, etc.) means fewer resources are available for users closer to the BS, who could otherwise achieve higher datarates with the same resources. That, combined with the increasing number of users in the dense urban area of Kathmandu, are considered to be the two main problems leading to such poor connectivity and, overall, poor data throughput. Therefore, our placement strategy targets offloading cellular users in low-coverage areas to nearby NNs.

\begin{algorithm}[t]

\caption{NN Placement}
\label{alg:placement}
{\footnotesize\begin{algorithmic}[1]
    \Output Returns list with NN placement: $\mathcal{P}$
    \Init
    \State Calculate $\max\mathrm{RSS}(Z), \quad \forall Z \in \mathcal{Z}$
    \State $\mathcal{P} \leftarrow \mathcal{C_\mathrm{LTE}}$
    \State $\mathcal{C_\mathrm{NN}} \leftarrow \mathcal{C} \smallsetminus \mathcal{C_\mathrm{LTE}}$
    \State $\text{define } d_\mathrm{p}, \quad d_\mathrm{p} \in \mathbb{N}^*$ \Comment{e.g., $d_\mathrm{p} = 100$}
    \State $\text{define } \max |\mathcal{P}| \in \mathbb{N}^*$ \Comment{e.g., $\max |\mathcal{P}| = 50$}

    \Level
    \Repeat
        \State  $\{ \triangle p_i p_j p_k \} \leftarrow \text{Delaunay}(\mathcal{P}), \quad \forall p \in \mathcal{P} \smallsetminus \{p_i, p_j, p_k\}, \, p \notin \text{Circumcircle}(\triangle p_i p_j p_k)$
        \State  $\text{IC} \leftarrow \mathcal{I}(\triangle p_i p_j p_k)$
        \State \text{Find} $\min d_\mathrm{P,IC_i}, \quad \forall P$

        \State $\text{IC} \leftarrow \text{IC}_i,  \quad  \text{with} \, \min d_\mathrm{P,IC_i} > d_\mathrm{p}, \quad \forall \text{IC}_i \in \text{IC}$
    
        \ForAll{$\text{IC}_i \in \text{IC}$}
            \State Calculate $\min d_\mathrm{C_{NN},IC_i}, \quad \forall C_\mathrm{NN} \in \mathcal{C_\mathrm{NN}}$
            \State $\text{pos}_i = \text{position corresponding to } \min d_\mathrm{C_{NN},IC_i}$
        \EndFor

        \ForAll{$\text{p}_i \in \text{pos}$}
            \State $d_\mathrm{Z,p_i} \leftarrow \text{calculate distance between } Z \text{ and } \text{p}_i, \quad \forall Z \in \mathcal{Z}$
            \State $Z_\mathrm{p_i} \leftarrow n \text{ nearest tiles, s.t. } d_\mathrm{Z,p_i}, \quad  n \in \mathbb{N}^*$ \Comment{e.g., $n = 1000$}
            \State $\mathrm{RSS}_{\mathrm{p_i}} = \sum\limits_{n=1}^{|\mathrm{Z_\mathrm{p_i}}|} \mathrm{RSS}(n) \bigg/  |\mathrm{Z_\mathrm{p_i}}|$
        \EndFor

        \State Calculate $\min\mathrm{RSS}_{p}, \quad \forall p \in \text{pos}$
        \State $\mathcal{P} \leftarrow \mathcal{P} \cup \{p\}, \quad \text{where } p \text{ the position with } \min\mathrm{RSS}_{p}$
        \State $\mathcal{C_\mathrm{NN}} = \mathcal{C_\mathrm{NN}} \smallsetminus p $
        \State Update $\max\mathrm{RSS}(Z), \quad \forall Z \in \mathcal{Z}$
    
    \Until{No more NNs can be added or $|\mathcal{P}| = \max |\mathcal{P}|$}

\end{algorithmic}}
\end{algorithm}

\subsection{NN Placement}
Based on the above, we devise a strategy for placing NNs across a map. Let $\mathcal{P} \triangleq \left \{1,\ldots,P \right \}$ denote the positions of all BSs on the map. We have $\mathcal{P}\subseteq\mathcal{C_\mathrm{LTE}}$ and $\mathcal{P}\subseteq\mathcal{C_\mathrm{NN}}$. We define a heuristic algorithm that prioritises NN placement at $P$ with low RSS, far away from existing LTE BSs (greater than a user-defined $d_\mathrm{p}$). Our algorithm is described in Alg.~\ref{alg:placement}.

Briefly, we start with all the existing LTE BSs (Fig.~\ref{fig:rssi_existing}) and iteratively add NNs to $\mathcal{P}$. We utilise Delaunay triangulation, forming triangular shapes between BSs in $\mathcal{P}$ (nodes) and the straight lines between them (edges). From all triangles $\{ \triangle p_i p_j p_k \}$, we find all in-centres ($\mathrm{IC}$) and exclude the ones close to an existing BS (comparing with $d_\mathrm{p}$). For the remaining $\mathrm{IC}$, we find the nearest building in-centre (in $\mathcal{C}_{\mathrm{NN}}$ that is considered a potential position $\mathrm{pos}$ for an NN placement. Using these potential positions, we find the $n$ nearest tiles on the road ($n$ is user-defined) and calculate their current average RSS. Out of all $|\mathrm{pos}|$ areas formed on the map, we identify the one with the lowest RSS and, consequently, the position for the next NN placement, i.e., the building in-centre we identified earlier. At every iteration that an NN is added, all RSS values are recalculated. Our algorithm iterates until no more NNs can be added or until a user-defined maximum number $\max |\mathcal{P}|$ is reached. Although this algorithm might lead to suboptimal results compared to an exhaustive search algorithm, it provides an effective and efficient placement of NNs in areas with low RSS (far away from existing BSs). The placed NNs are intended to accommodate users in low coverage areas that introduce delays and underutilisation of resources, increasing the system's performance overall. 

\subsection{Concurrent Users and Datarate Calculations}
In Sec.~\ref{subsec:datarate_adaptation}, we defined $D_\mathrm{MCS}$ in accordance to the $P_\mathrm{rx}$, the sensitivity levels, the MCS, and the number of cell sectors. The estimated datarate is achieved within a given $BW_\mathrm{c}$, considering a single channel. Assuming that an entire LTE band $f_{\text{b}} = [f_{\text{b, min}}, f_{\text{b, max}}]$ (in \si{\hertz}) is available (considering e.g., the downlink), the total concurrent users a BS can serve (without reduced datarate) becomes $u_\mathrm{RAT} = (\nicefrac{360^{\circ}}{\theta}) \, \frac{f_{\text{b, min}} - f_{\text{b, min}}}{BW_\mathrm{c}}$. 

The perceived RSS values indicate the tiles or buildings each BS services, thus, by extension the users within it. For a given $Z$ or $B$, we identify the BSs that can provide connectivity ($P_\mathrm{rx}$ higher than the lowest sensitivity level). The BS with the highest RSS is selected to serve the given $Z$ or $B$. As the existing LTE infrastructure and the NNs introduced are fairly similar, we assume if multiple technologies serve the same tile, a user will always prefer the one with the highest RSS.

At this point, we introduce the concept of active users within our system. This models \emph{RRC-Connected} and \emph{RRC-Idle} states found in LTE and can be expressed as a percentage of the total users. Therefore, the number of connected users to a BS is $u_{\mathrm{BS}} = \sum_{i=1}^{Z_{\mathrm{BS}}} \beta \, u_{Z,i} + \sum_{i=1}^{B_{\mathrm{BS}}} \beta \, u_{\mathrm{B},i}$ where $\beta$ is the probability for a user to be active, $Z_{\mathrm{BS}}$ being all tiles and $B_{\mathrm{BS}}$ all the buildings served by a given BS.

If $u_{\mathrm{BS}} \leq u_\mathrm{RAT}$, then the average datarate for a given BS is $D_\mathrm{BS}^* = \nicefrac{1}{u_{\mathrm{BS}}} \sum_{i=1}^{u_{\mathrm{BS}}} D_{\mathrm{MCS,i}}u_{\mathrm{BS}}$. If $u_{\mathrm{BS}} > u_\mathrm{RAT}$, a simple Time Division Multiple Access (TDMA) is considered and all users are allocated the same time over the channel. Therefore, our datarate becomes $D_\mathrm{BS}^* = \nicefrac{1}{u_{\mathrm{BS}}} \sum_{i=1}^{u_{\mathrm{BS}}} D_{\mathrm{MCS,i}} \, \nicefrac{u_\mathrm{RAT}}{u_{\mathrm{BS}}}$.

\begin{table}[t]
\renewcommand{\arraystretch}{1}
    \centering
    \caption{List of Simulation Parameters.}
    \begin{tabular}{rl|c|c}
        \multicolumn{4}{c}{LTE BSs and NNs} \\
     \toprule
        \multicolumn{2}{c}{Parameter} & Provider BS & Nomadic Node \\
       \midrule
         LTE Band & $f_\mathrm{b}$ & \multicolumn{2}{c}{\SIrange{1805}{1880}{\mega\hertz}} \\
         Transmission Power & $P_\mathrm{tx}$ &  \SI{35}{dBm}& \SI{30}{dBm}\\
         TX Antenna Gain & $G_\mathrm{tx}$ & \SI{15}{dBi}& \SI{16}{dBi}\\
         RX Antenna Gain & $G_\mathrm{rx}$ & \multicolumn{2}{c}{\SI{0}{dBi}} \\
         Carrier Frequency & $f_\mathrm{c}$ & \SI{1850}{MHz}&\SI{1850}{MHz}\\
         Bandwidth & $BW_\mathrm{c}$ & \SI{15}{MHz} & \SI{20}{MHz}\\
         BS Height & $h_\mathrm{BS}$ & \SIrange{5}{15}{\meter} & adjustable\\
         Beamwidth Angle & $\theta$ & $120^\circ$ & 60$^\circ$ \\
         LTE Overhead & $O$ & \multicolumn{2}{c}{\SI{25}{\%}} \\
     \bottomrule   
        \multicolumn{4}{c}{} \\
        \multicolumn{4}{c}{Other Parameters} \\
     \toprule     
        \multicolumn{2}{c}{Parameter} & \multicolumn{2}{c}{Value} \\
       \midrule
       Distance & $d_\mathrm{p}$ & \multicolumn{2}{c}{\SI{100}{\meter}} \\
       Tile Size & $Z$ & \multicolumn{2}{c}{$\SI{4}{} \times \SI{4}{\meter}^2$} \\
       Active Users Prob. & $\beta$ & \multicolumn{2}{c}{0.25} \\
       No. of tiles & $n$ & \multicolumn{2}{c}{1000} \\
     \bottomrule   
    \end{tabular}

    \label{tab:Specs}
\end{table}

\section{Results and Discussion}
\label{sec:Results}
Our scenario consists of $50$ ``simulated hours'', starting at $12$~a.m., and is split in $50$ equal timesteps. For each timestep, we generate the user mobility traces and the building users based on the above-described models. We evaluate the RSS and the perceived datarate for different numbers of NNs. Table~\ref{tab:Specs} summarises the simulation parameters. We finally considered a SISO setup for both LTE and NN deployments.

\subsection{RSS Heatmaps and CDFs}
\begin{figure}
   \centering
    \includegraphics[width=\columnwidth]{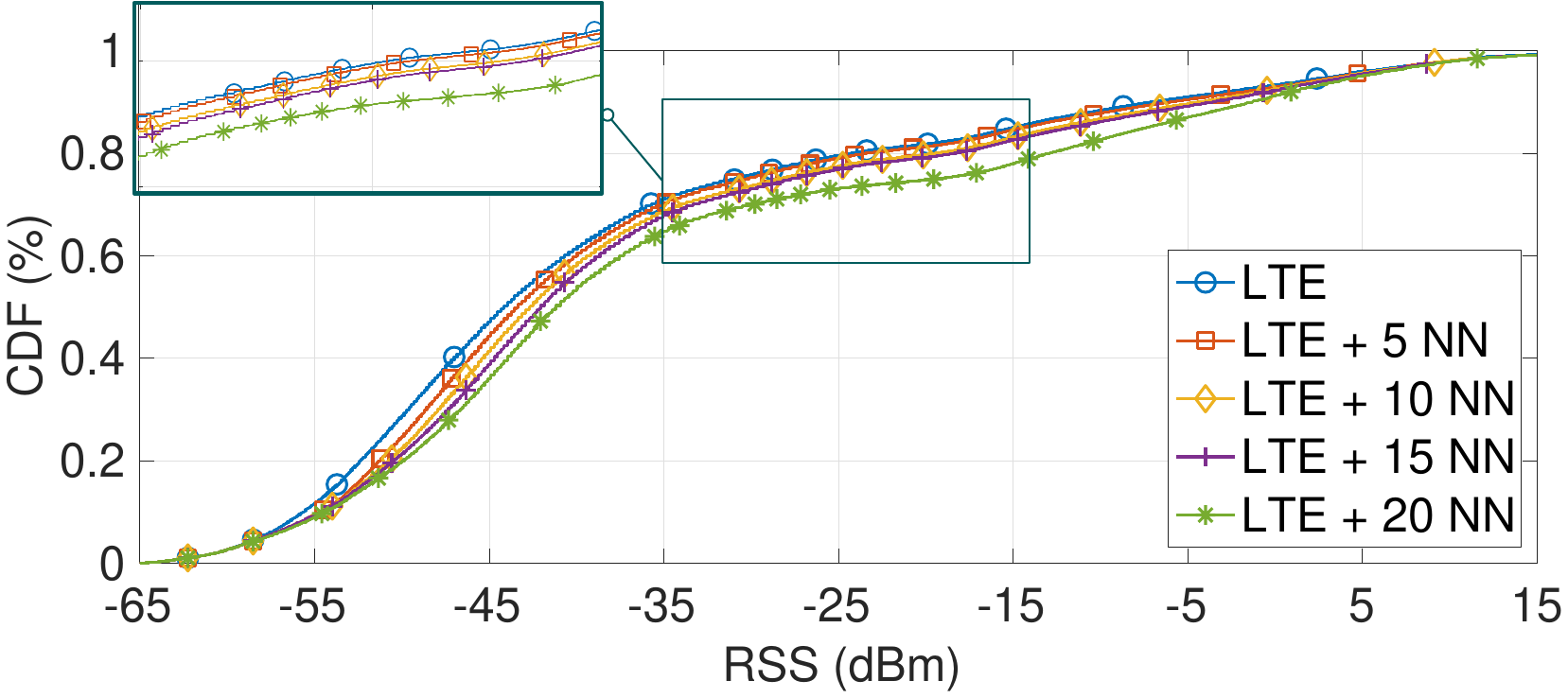}
    \vspace{-6mm}
   \caption{RSS for LTE and different numbers of NN placements.}
   \label{fig:rssi}
\end{figure}

In Fig.~\ref{fig:rssi}, we present the RSS results per tile for LTE and the optimal placement of a different number of NNs, from $5$ to $20$, with a step of $5$. As NNs are cheaper than a macrocell LTE BS, and with our scenario starting with $49$ pre-existing LTE BSs, $20$ NNs are considered for deployment, requiring a small initial capital investment. After deployment, we observe a transitional shift to the right as the number of NNs increases, which indicates an improvement in the network coverage. It is observed that the main area of improvement lies between the \SIrange{-55}{0}{dBm} values. This can be explained by the fact that our NN placement intervention excludes the areas around the edges of the cropped map, keeping some areas unaltered in the lower and upper RSS range. The effectiveness of the NNs on the improved RSS is evident from the low ranges even when $5$ NNs are placed. The shifting increases with more NNs, with significant gains observed by adding 20 NNs. $50\%$ of the RSS values fall below \SI{-45}{dBm} for LTE (median), while with the addition of the NNs, this value shifts to \SI{40}{dBm} when we place $20$ NNs. To further investigate the effectiveness of the NN intervention, we run two-sample non-parametric Kolmogorov-Smirnov statistical tests~\cite{walpole2017probability} between the LTE (baseline) and the combination with the NNs distributions. All four tests resulted in $p-$values lower than \SI{0.05}{} and confirmed the statistical significance of the improved RSS distributions.

\subsection{Average Datarate Distributions}
\begin{figure}
   \centering
    \includegraphics[width=\columnwidth]{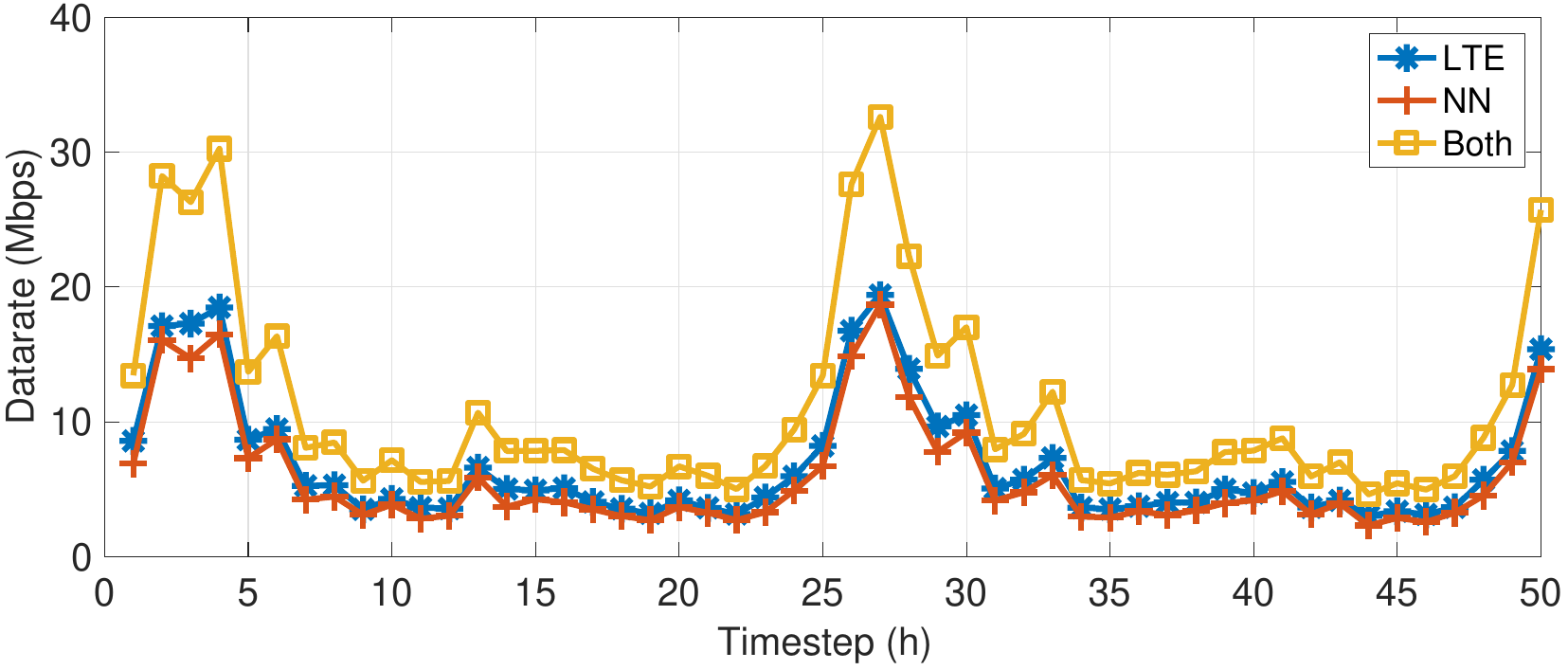}
    \vspace{-6mm}
   \caption{Averaged datarate for LTE, NN, and both technologies for all timesteps (considering 20 NNs).}
   \label{fig:datarate}
\end{figure}

\begin{figure}[t]
    \centering
    \begin{subfigure}[b]{\columnwidth}
        \centering
        \includegraphics[width=\columnwidth]{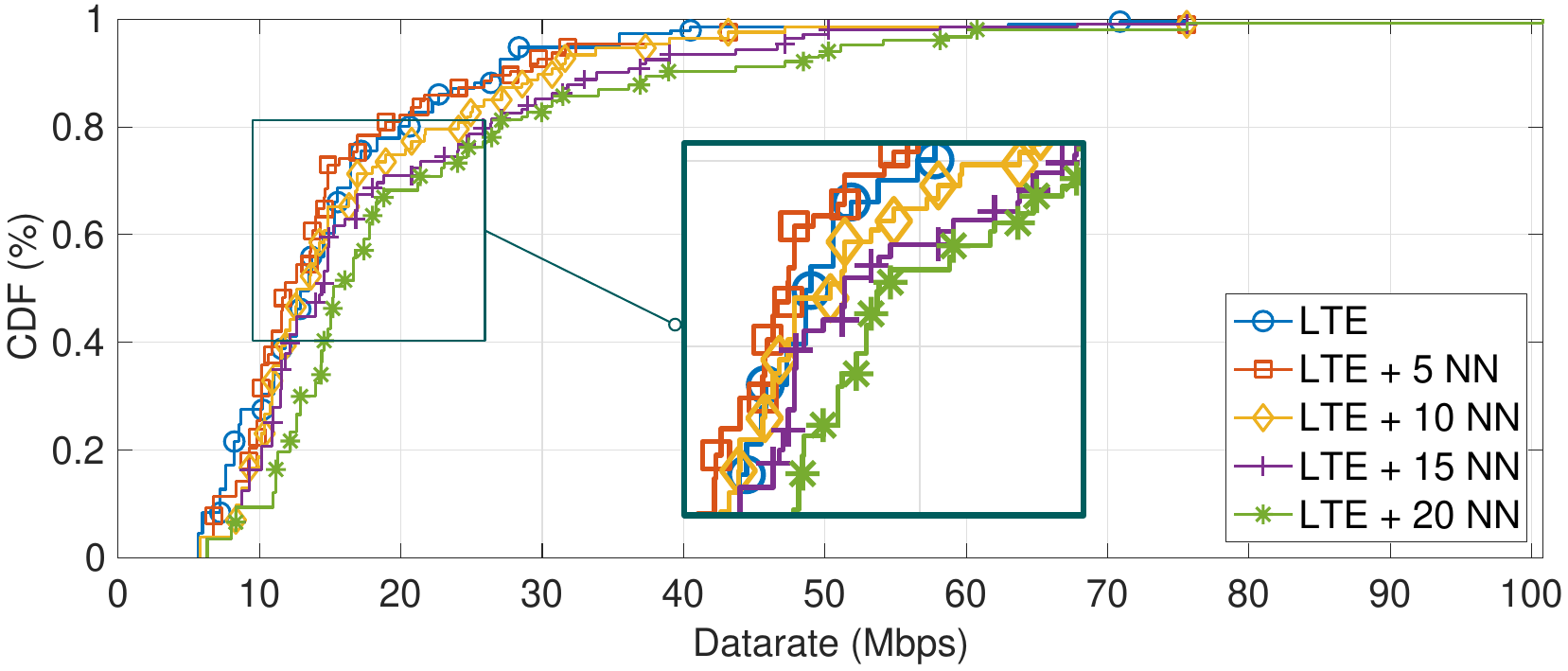}
        \vspace{-6mm}
        \caption{Timestep no. 19 - highest user density.}
        \label{fig:timestep19}
    \end{subfigure}
    
    \vspace{2mm} 

    \begin{subfigure}[b]{\columnwidth}
        \centering
        \includegraphics[width=\columnwidth]{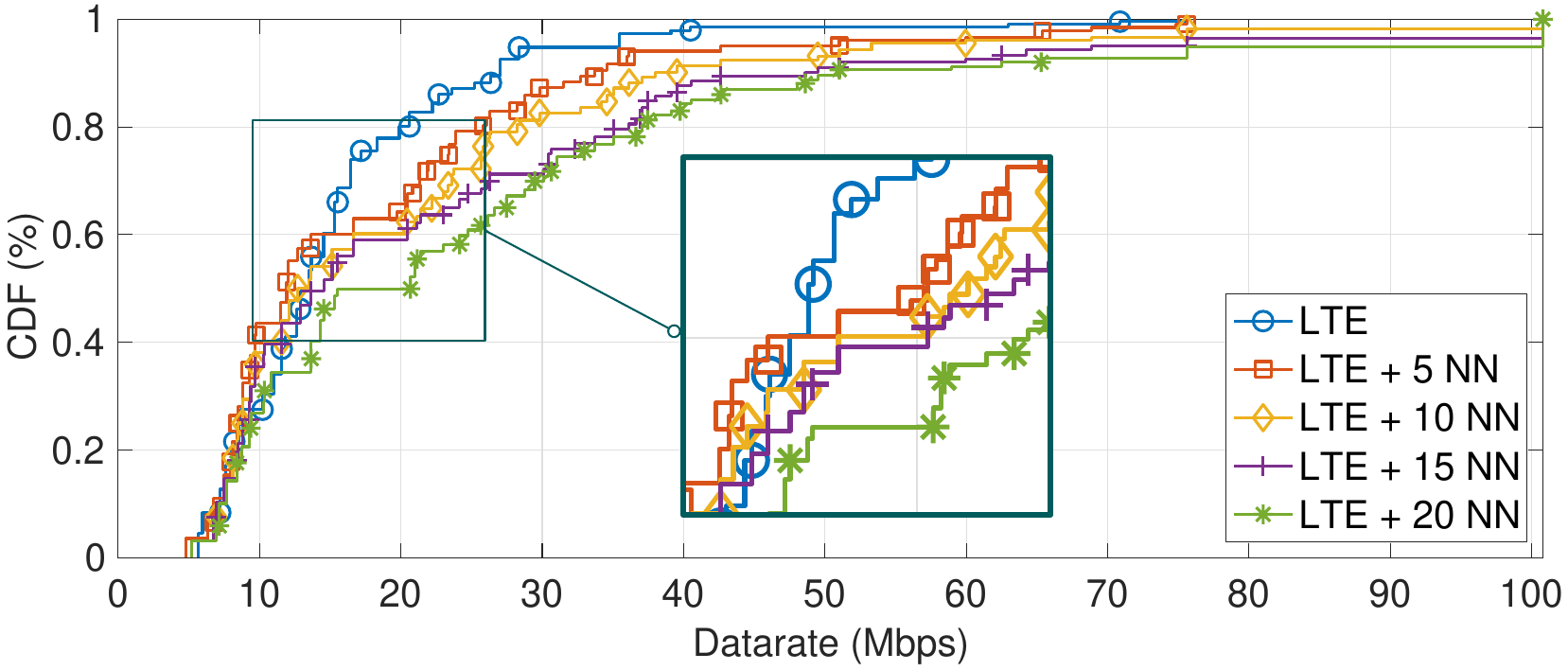}
        \vspace{-6mm}
        \caption{Timestep no. 27 - lowest user density.}
        \label{fig:timestep27}
    \end{subfigure}
    \vspace{-3mm}
    \caption{CDF plots for the highest and lowest user densities across the scenario, analysed for different no. of NNs.}
    \label{fig:main}
\end{figure}

\begin{figure}
   \centering
    \includegraphics[width=\columnwidth]{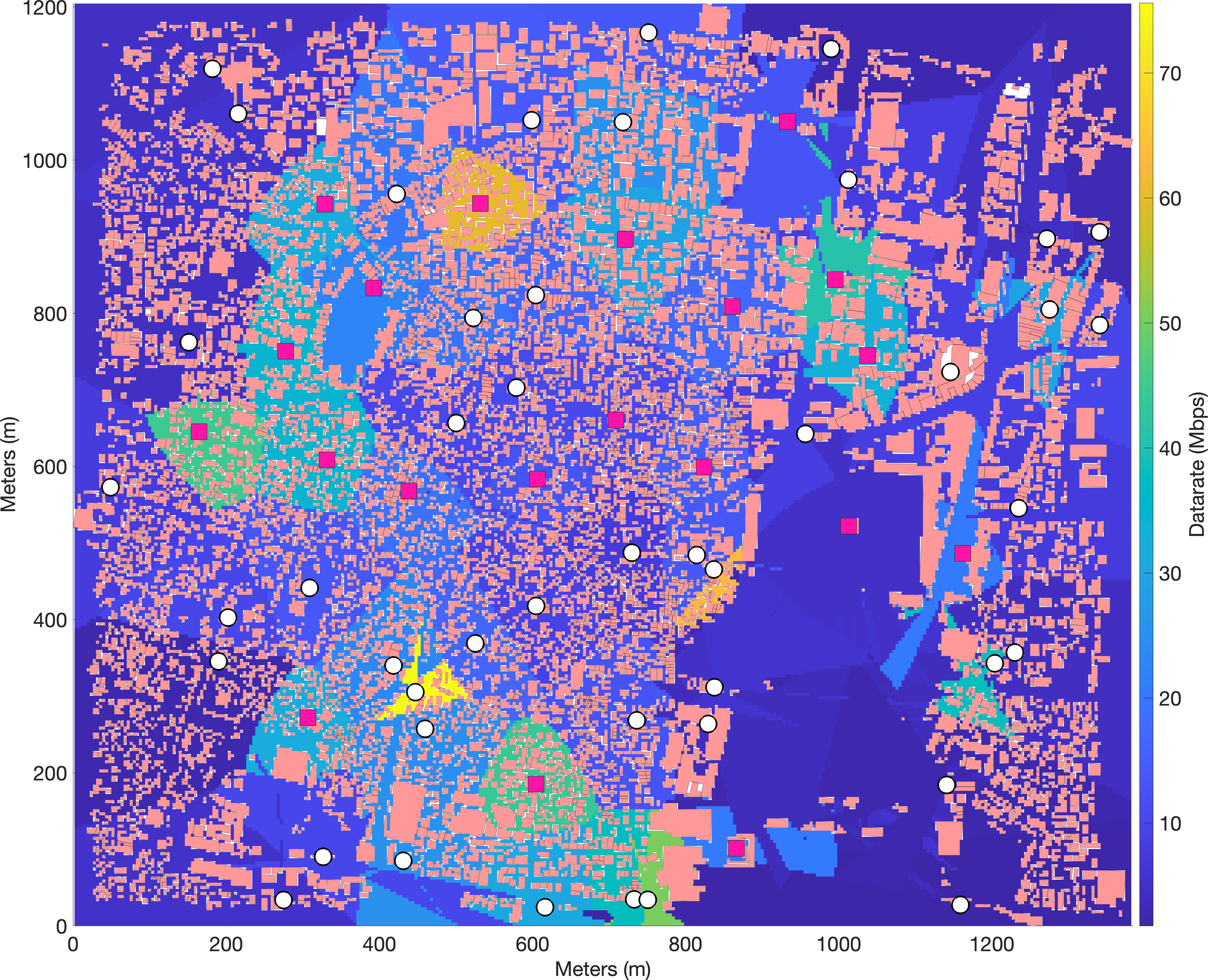}
   \caption{Example datarate heatmap for both technologies at timestep no. 10 - the white circles denote the LTE BSs and the red squares the NNs.}
   \label{fig:datarateHeatmap}
\end{figure}

The effectiveness of the NNs placement is reflected by the datarate summarised in Figs.~\ref{fig:datarate}-\ref{fig:datarateHeatmap}. The average datarate of the whole area for the LTE, the NN, and the combined solution (with $20$ NNs) over the different timesteps is illustrated in Fig.~\ref{fig:datarate}. From this figure, it can be observed that all three solutions follow the same pattern with different magnitudes. Namely, the average datarate is peaking in the early hours, when the number of users decreases in the area of interest while dropping, as anticipated, around rush hour in the morning and through the day when the number of active users increases. Overall, we observe an increase in the datarate by around $65\%$. LTE and NN individually achieve \SI{6.79}{\Mbps} and \SI{5.86}{\Mbps}, respectively, while the combination boosts the performance to \SI{11.18}{\Mbps}.

Furthermore, we investigate the CDFs of the datarates at the different time steps of the simulation; at timestep $19$ that represents ``peak user density'' and at timestep $27$ that represents an ``off-peak'' scenario. Although the CDF curves of the LTE and the combination with the NNs are interlacing for a small number of NNs, it can be easily observed that in both cases, the addition of $20$ NNs significantly improves the median average datarate; about $50\%$ for timestep $19$ and about $60\%$ for timestep $27$ when considering a datarate of \SI{20}{\Mbps} (an average datarate required for a 4K HDR video streaming experience). 

An example heatmap for a peak time at timestep $10$ is also provided in Fig.~\ref{fig:datarateHeatmap} to demonstrate the datarate distribution across the entire map. As observed, the NNs (red squares) are optimally positioned to cover the areas between the existing BSs (white circles), increasing the datarate around the areas they cover. Moreover, offloading the users to the newly placed NNs also increases the LTE's datarate (e.g., we see a few LTE BSs with bright yellow colours around them). Overall, the above NN placement strategy and the decision to offload users with poor LTE connectivity to the newly introduced plane are proven to have significant benefits on the overall network performance, even with a small number of utilised NNs and achieving our initial aim for a cost-effective, and efficient optimisation. 

\section{Conclusion}
\label{sec:Conclusions}
In this paper, we investigated the feasibility of deploying NNs to improve cellular connectivity in the dense urban environment of Kathmandu, Nepal. We investigate our scenario within a DT environment using the DRIVE framework. Exploring the impact of NN placement on network performance, we find that NNs can effectively bridge connectivity gaps and manage traffic loads in heavily populated areas, providing a practical solution to the challenges posed by urbanisation and limited infrastructure investment. This study highlights the potential of NNs not only for enhancing connectivity in urban settings but also for their adaptability in dynamic and high-demand scenarios. Future work will focus on refining the DT models, investigating more advanced placement solutions and cell-switching strategies according to user mobility, and finally exploring the deployment of NNs in combination with other non-cellular technologies.

\section*{Acknowledgment}

This work is partly funded by the Royal Academy of Engineering (RAEng) Frontiers Seedcorn Funding FS-2223-19-13 and by Research England ODAISPF24-11. This work is also a contribution by Project REASON, a UK Government funded project under the Future Open Networks Research Challenge (FONRC) sponsored by the Department of Science Innovation and Technology (DSIT).

The authors would like to thank the RAEng collaborators Sushmita Shrestha and Doma Tsering Tamang from Utopia Nepal, Prof. Jyotsna Bapat and Dr Dibakar Das from IIIT Bangalore for the insightful discussions.

\bibliographystyle{IEEEtran}
\bibliography{refs}

\begin{thebibliography}{10}
\providecommand{\url}[1]{#1}
\csname url@samestyle\endcsname
\providecommand{\newblock}{\relax}
\providecommand{\bibinfo}[2]{#2}
\providecommand{\BIBentrySTDinterwordspacing}{\spaceskip=0pt\relax}
\providecommand{\BIBentryALTinterwordstretchfactor}{4}
\providecommand{\BIBentryALTinterwordspacing}{\spaceskip=\fontdimen2\font plus
\BIBentryALTinterwordstretchfactor\fontdimen3\font minus \fontdimen4\font\relax}
\providecommand{\BIBforeignlanguage}[2]{{%
\expandafter\ifx\csname l@#1\endcsname\relax
\typeout{** WARNING: IEEEtran.bst: No hyphenation pattern has been}%
\typeout{** loaded for the language `#1'. Using the pattern for}%
\typeout{** the default language instead.}%
\else
\language=\csname l@#1\endcsname
\fi
#2}}
\providecommand{\BIBdecl}{\relax}
\BIBdecl

\bibitem{un17sdg}
\BIBentryALTinterwordspacing
``{United Nations: 17 Sustainable Development Goals},'' 2022, {Accessed: 12/09/2024}. [Online]. Available: \url{https://sdgs.un.org/goals#goals}
\BIBentrySTDinterwordspacing

\bibitem{pelton:2019}
J.~Pelton and I.~Singh, \emph{{Smart Cities of Today and Tomorrow}}.\hskip 1em plus 0.5em minus 0.4em\relax Springer International Publishing, 2019.

\bibitem{yaacoub:2020}
E.~Yaacoub and M.~S. Alouini, ``{A Key 6G Challenge and Opportunity—Connecting the Base of the Pyramid: A Survey on Rural Connectivity},'' \emph{Proc. of the IEEE}, vol. 108, no.~4, pp. 533--582, Mar. 2020.

\bibitem{5GUseCases}
O.~O. Erunkulu \emph{et~al.}, ``{5G Mobile Communication Applications: A Survey and Comparison of Use Cases},'' \emph{IEEE Access}, vol.~9, pp. 97\,251--97\,295, 2021.

\bibitem{MeasuringDigitalDevelopment_ITU2023}
\BIBentryALTinterwordspacing
``{Measuring Digital Development: Facts and Figures 2023},'' {International Telecommunication Union (ITU)}, 2023, {Accessed: 12/09/2024}. [Online]. Available: \url{https://www.itu.int/en/ITU-D/Statistics/Pages/facts/default.aspx}
\BIBentrySTDinterwordspacing

\bibitem{subramanian2006rethinking}
L.~Subramanian \emph{et~al.}, ``{Rethinking Wireless for the Developing World},'' \emph{Irvine Is Burning}, vol.~43, 2006.

\bibitem{Ericsson2023}
\BIBentryALTinterwordspacing
Ericsson, ``Ericsson mobility report june 2023,'' 2023, {Accessed: 12/09/2024}. [Online]. Available: \url{https://www.ericsson.com/en/reports-and-papers/mobility-report/reports/june-2023}
\BIBentrySTDinterwordspacing

\bibitem{GSMA2022}
\BIBentryALTinterwordspacing
GSMA, ``The mobile economy 2022,'' 2022, {Accessed: 12/09/2024}. [Online]. Available: \url{https://www.gsma.com/mobileeconomy/}
\BIBentrySTDinterwordspacing

\bibitem{dang2020should}
S.~Dang \emph{et~al.}, ``{What should 6G be?}'' \emph{Nature Electronics}, vol.~3, no.~1, pp. 20--29, 2020.

\bibitem{procBulakci}
O.~Bulakci \emph{et~al.}, ``{Towards Flexible Network Deployment in 5G: Nomadic Node Enhancement to Heterogeneous Networks},'' in \emph{Proc. of IEEE ICCW}, Jun. 2015, pp. 2572--2577.

\bibitem{driveSimulator}
I.~{Mavromatis} \emph{et~al.}, ``{DRIVE: A Digital Network Oracle for Cooperative Intelligent Transportation Systems},'' in \emph{Proc. of IEEE ISCC}, Jul. 2020.

\bibitem{DriveGithub}
I.~Mavromatis, ``{DRIVE: Digital twin for self-dRiving Intelligent VEhicles},'' \url{https://github.com/ioannismavromatis/DRIVE_Simulator}, {Accessed: 12/09/2024}.

\bibitem{OpenStreetMaps}
O.~Foundation, ``{OpenStreetMap},'' \url{https://www.openstreetmap.org/about}, {Accessed: 12/09/2024}.

\bibitem{CellMapper}
CellMapper, ``Cellmapper: Mobile network coverage mapping,'' \url{https://www.cellmapper.net}, {Accessed: 12/09/2024}.

\bibitem{sym15010002}
V.~Stoynov \emph{et~al.}, ``{Ultra-Dense Networks: Taxonomy and Key Performance Indicators},'' \emph{Symmetry}, vol.~15, no.~1, 2023.

\bibitem{mmwaveInfrastructureITS}
I.~{Mavromatis} \emph{et~al.}, ``{Efficient Millimeter-Wave Infrastructure Placement for City-Scale ITS},'' in \emph{Proc. of IEEE VTC2019-Spring}, Apr. 2019, pp. 1--5.

\bibitem{Asad_SurveyMobility}
S.~M.~A. Zaidi \emph{et~al.}, ``{Mobility Management in Emerging Ultra-Dense Cellular Networks: A Survey, Outlook, and Future Research Directions},'' \emph{IEEE Access}, vol.~8, pp. 183\,505--183\,533, 2020.

\bibitem{costOptimalMmWaves}
M.~Dong \emph{et~al.}, ``{Cost-Optimal Deployment of Millimeter-Wave Base Stations Under Outage Requirement},'' \emph{{IEEE Trans. Wirel. Comms.}}, vol.~21, no.~12, pp. 10\,544--10\,559, 2022.

\bibitem{MOLLAHASANI2020107271}
S.~Mollahasani \emph{et~al.}, ``{Density-aware mobile networks: Opportunities and challenges},'' \emph{Computer Networks}, vol. 175, p. 107271, 2020.

\bibitem{6214336}
P.~Kolios, V.~Friderikos, and K.~Papadaki, ``{Mechanical Forwarding for Nomadic Mobility in Cellular Networks},'' in \emph{Proc. of IEEE WCNC}, 2012, pp. 3091--3096.

\bibitem{8108598}
T.~Sahin \emph{et~al.}, ``{Flexible Network Deployment in 5G: Performance of Vehicular Nomadic Nodes},'' in \emph{Proc. of IEEE VTC Spring}, Jun. 2017, pp. 1--6.

\bibitem{dt6G}
N.~P. Kuruvatti \emph{et~al.}, ``{Empowering 6G Communication Systems With Digital Twin Technology: A Comprehensive Survey},'' \emph{IEEE Access}, vol.~10, pp. 112\,158--112\,186, 2022.

\bibitem{Liu2019}
L.~Liu \emph{et~al.}, ``{Time-domain ICIC and Optimized Designs for 5G and Beyond: A Survey},'' \emph{Science China Information Sciences}, vol.~62, no.~2, p. 21302, 2019.

\bibitem{book_Zhang_2024}
Z.~Yan, \emph{{Digital Twin Architectures, Networks, and Applications}}.\hskip 1em plus 0.5em minus 0.4em\relax Springer Nature, 2024.

\bibitem{Dahal_2019}
R.~K. Dahal, ``{Customer Satisfaction in Nepalese Cellular Networks},'' \emph{Tribhuvan University Journal}, vol.~33, no.~2, p. 59–72, Dec. 2019.

\bibitem{polygonUnion}
F.~Martínez, A.~J. Rueda, and F.~R. Feito, ``{A New Algorithm for Computing Boolean Operations on Polygons},'' \emph{Computers \& Geosciences}, vol.~35, no.~6, pp. 1177--1185, Aug. 2009.

\bibitem{3GPP38901}
{3rd Generation Partnership Project (3GPP)}, ``{Study on Channel Model for Frequencies from 0.5 to 100 GHz},'' 3GPP, Tech. Rep. TR 38.901, Apr. 2022, release 17, V17.0.0.

\bibitem{3gpp-ts-36-101-v17-6-0}
------, ``{3GPP TS 36.101},'' 3GPP, Technical Specification TS 36.101, 2023, release 17, V17.6.0.

\bibitem{etsi_ts_136211}
{European Telecommunications Standards Institute (ETSI)}, ``Physical channels and modulation,'' ETSI, Technical Specification TS 136 211, 2022, release 17, V17.1.0.

\bibitem{spatialModelling}
D.~Lee \emph{et~al.}, ``{Spatial modeling of the traffic density in cellular networks},'' \emph{IEEE Wirel. Commun.}, vol.~21, no.~1, pp. 80--88, Feb. 2014.

\bibitem{temporalModelling}
S.~Wang \emph{et~al.}, ``{An Approach for Spatial-Temporal Traffic Modeling in Mobile Cellular Networks},'' in \emph{Proc. of Int. Telet Congr.}, Sep. 2015, pp. 203--209.

\bibitem{timeAlignment}
X.~Wang \emph{et~al.}, ``{Spatio-temporal analysis and prediction of cellular traffic in metropolis},'' in \emph{Proc. of IEEE ICNP}, Oct. 2017, pp. 1--10.

\bibitem{jiang2024spatiotemporal}
D.~Jiang \emph{et~al.}, ``{A Spatiotemporal Hierarchical Analysis Method for Urban Traffic Congestion Optimization Based on Calculation of Road Carrying Capacity in Spatial Grids},'' \emph{ISPRS Int. J. of Geo-Information}, vol.~13, no.~2, p.~59, Dec. 2024.

\bibitem{cesario2024detecting}
E.~Cesario \emph{et~al.}, ``{Discovering Multi-density Urban Hotspots in a Smart City},'' in \emph{Proc. of IEEE SMARTCOMP}, Sep. 2020, pp. 332--337.

\bibitem{mATRIC}
REASON-Open-Networks, ``Reason: Realising enabling architectures and solutions for open networks,'' \url{https://reason-open-networks.ac.uk/about/matric/}, {Accessed: 12/09/2024}.

\bibitem{5g-victori:2021}
L.~Bassbouss \emph{et~al.}, ``{5G-VICTORI: Optimizing Media Streaming in Mobile Environments Using mmWave, NBMP and 5G Edge Computing},'' \emph{AIAI IFIP WG 12.5}, 2021.

\bibitem{walpole2017probability}
R.~E. Walpole \emph{et~al.}, \emph{{Probability and Statistics for Engineers and Scientists}}, 9th~ed.\hskip 1em plus 0.5em minus 0.4em\relax Pearson, 2017.

\end{thebibliography}

\end{document}